\documentclass[aps, prd, reprint]{revtex4-2} 
\usepackage{graphicx,amsmath,amssymb,epsf} \usepackage{latexsym,bm,
  slashed} \usepackage{xcolor} \definecolor{dark}{rgb}{0.10,0.2,0.3}
\definecolor{magenta}{rgb}{0.7,0.1,0.3}
\definecolor{purpure}{rgb}{0.5,0.15,0.3}
\usepackage[font=small,format=plain,labelfont=bf,up,textfont=it,up]{caption}
\usepackage{hyperref} \hypersetup{colorlinks, citecolor=blue,
  filecolor=blue, linkcolor=magenta,
  urlcolor=purpure,hyperfootnotes=true,pdftex} 
\newcommand{\tr}{{\rm tr}}
\newcommand{\kt}{{\bm k}}
\newcommand{\pt}{{\bm p}}
\newcommand{\qt}{{\bm q}}
\newcommand{\lt}{{\bm l}}
\newcommand{\xit}{{\bm \xi}}

\newcommand{\bt}{{\bm b}}

\begin{document}

\title{
Transverse Momentum Dependent Gluon Distribution within High Energy Factorization  at Next-to-Leading Order
   } \author{\large Martin~Hentschinski}
\affiliation{
 Departamento de Actuaria, F\'isica y Matem\'aticas, 
Universidad de las Americas Puebla, Ex-Hacienda Santa Catarina Martir S/N, San Andrés Cholula,  72820 Puebla, Mexico 
}
\email{martin.hentschinski@udlap.mx}

\begin{abstract}
  We discuss Transverse Momentum Dependent (TMD) gluon distributions
  within high energy factorization at next-to-leading order in the
  strong coupling within the framework of Lipatov's high energy
  effective action. We provide a detailed discussion of both rapidity
  divergences related to the TMD definition and its soft factor on the
  one hand, and rapidity divergences due to high energy factorization
  on the other hand, and discuss common features and differences
  between Collins-Soper (CS) and Balitsky-Fadin-Kuarev-Lipatov (BFKL)
  evolution. While we confirm earlier results which state that the
  unpolarized and linearly polarized gluon TMD agree in the BFKL limit
  at leading order, we find that both distributions differ, once
  next-to-leading order corrections are being included. Unlike
  previous results, our framework allows to recover the complete
  anomalous dimension associated with Collins-Soper-Sterman (CSS)
  evolution of the TMD distribution, including also single-logarithmic
  terms in the CSS evolution. As an additional result we provide a
  definition of $k_T$-factorization, {\it i.e.} matching of off-shell
  coefficients to collinear factorization at next-to-leading order
  within high energy factorization and the effective action
  framework. We furthermore establish a link between the QCD operator
  definition of the TMD gluon distribution and a previously derived
  off-shell TMD gluon-to-gluon splitting function, which is within the
  present framework obtained as the real 1-loop correction.
\end{abstract}
\maketitle

\section{Introduction}
\label{sec:intro}

Transverse momentum dependent (TMD) parton distribution functions
(PDFs)~\cite{Angeles-Martinez:2015sea,Collins:2011zzd, Abdulov:2021ivr} are objects of increased
interest, since they allow to provide a more precise kinematic
description of partonic scattering processes already at the leading
order (LO) of perturbation theory. This is in particular true, if the
observable of interest is not entirely inclusive. In that case, TMD
PDFs provide an important advantage over a description based on
collinear parton distributions.  TMD PDFs arise naturally in processes
which are characterized by a hierarchy of scales. With $Q$ the scale of
the hard reactions, TMD PDFs are at first defined for the hierarchy
$Q \gg q_T \gg \Lambda_{\text{QCD}}$ with $q_T$ the transverse
momentum of the parton and $\Lambda_{\text{QCD}}$ the QCD
characteristic scale of a few hundred MeV. The QCD description of such
events gives then rise to the so-called Collins-Soper-Sterman (CSS)
\cite{Collins:1981uw,Collins:1981uk,Collins:1984kg} resummation
formalism. A different kinematic hierarchy in which TMD PDFs arise is
provided by the perturbative Regge or low $x$ limit
$\sqrt{s} \gg M \gg \Lambda_{\text{QCD}}$, where $\sqrt{s}$ denotes
the center of mass energy of the reaction and $x = M^2/s$. While the
resulting high energy factorization 
\cite{Catani:1990xk,Catani:1990eg,Catani:1993ww}
does not primarily address the
description of transverse momenta of final states, the ensuing
formalism naturally factorizes cross-sections  into transverse momentum dependent coefficients, so
called impact factors, and  transverse momentum dependent Green's function, which summerize logarithms in the center of mass energy. In particular 
 Balitsky-Fadin-Kuraev-Lipatov
(BFKL) evolution  \cite{Kuraev:1976ge, Lipatov:1976zz,
  Kuraev:1977fs, Balitsky:1978ic, Fadin:1998py, Ciafaloni:1998gs}, as well as its non-linear extensions, 
keep  track of transverse momenta along the evolution chain.\\

Both kinematic limits have a region of overlap, characterized through
the hierarchy $\sqrt{s} \gg M \gg q_T \gg \Lambda_{\text{QCD}}$, which
is of particular interest due its sensitivity to the emergence of a
semi-hard dynamical scale in the low $x$ limit, the so-called
saturation scale \cite{Gribov:1984tu}.  The relation of both
frameworks has been explored in a series of publications
\cite{Dominguez:2011wm, Dominguez:2011br, Balitsky:2015qba,
  Zhou:2016tfe, Marquet:2016cgx, Altinoluk:2019fui, Altinoluk:2021ygv,
  Xiao:2017yya, Nefedov:2021vvy} and is currently used for a wide set
of phenomenological studies see {\it e.g.}\cite{Dumitru:2018kuw,
  Stasto:2018rci, Mantysaari:2019hkq,  vanHameren:2020rqt, Fujii:2020bkl, Boussarie:2021lkb}. A somehow
orthogonal approach has been put forward in
\cite{Gituliar:2015agu,Hentschinski:2016wya,Hentschinski:2017ayz}:
instead of studying the region of overlap of both kinematic regimes,
the goal has been to derive TMD evolution kernels which are meant to
achieve a simultaneous resummation of both
Dokshitzer-Gribov-Lipatov-Altarelli-Parisi (DGLAP) and BFKL
logarithms \cite{Catani:1994sq, Catani:1993rn}. Such an approach seems to be of particular interest for
Monte-Carlo applications such as \cite{Jung:2010si, Hautmann:2012sh, Hautmann:2019biw, BermudezMartinez:2019anj, Baranov:2021uol}.  While
currently only real splitting kernels have been derived, which reduce
in the regarding limits to the respective real DGLAP and BFKL kernels, the relation to
CSS resummation is at the moment less clear, however the gluon-to-gluon
splitting kernel could be shown to reduce to the
Ciafaloni-Catani-Fiore-Marchesini kernel \cite{Ciafaloni:1987ur,
  Catani:1989yc} in the soft limit.
\\

In the following we aim at solving various open questions related to
the previously mentioned studies. In particular we will give a
next-to-leading order (NLO) study with respect to an expansion in the
strong coupling constant $\alpha_s$ of the gluon TMD in the high
energy limit. While a related study has been already presented in
\cite{Xiao:2017yya} within the Color-Glass-Condensate approach, we
will investigate this problem within the context of Lipatov's high
energy effective action \cite{Lipatov:1995pn,Lipatov:1996ts}. Starting
with \cite{Hentschinski:2011tz,Hentschinski:2011xg,Chachamis:2012cc,
  Chachamis:2012gh, Chachamis:2013hma} the systematic determination of
perturbative higher order corrections has been worked out for this
framework, while in \cite{Hentschinski:2018rrf} equivalence with the
Color-Glass-Condensate formalism has been demonstrated, including an
re-derivation of the Balitsky-JIMWLK evolution, see also
\cite{Chachamis:2012mw, Hentschinski:2020rfx} for reviews; for further
recent studies based on this framework see also \cite{Kotko:2017nkx,
  Bondarenko:2017vfc,vanHameren:2017hxx,Bondarenko:2018kqs,Braun:2020thc,Blanco:2020akb,GomezBock:2020zxp,Bondarenko:2021rbp}.
In \cite{Hentschinski:2020tbi} a systematic framework for the
determination of next-to-leading order corrections at cross-section
level has been worked out. For the present study we will further
extend this framework to include asymmetric factorization scale
settings as required for a matching to collinear factorization, {\it
  i.e.} $k_T$-factorization \cite{Catani:1990eg}. While this framework
is currently limited to the dilute regime, {\it i.e.} 2 (reggeized)
gluon exchange at the level of the cross-section, it has the great
advantage that it allows to systematically study different choices of
factorization parameter and schemes, which will be of particular use
for the further exploration of the relation between BFKL and Collins-Soper
(CS) \cite{Collins:1981uw,Collins:1981uk} evolution, initiated in
\cite{Zhou:2016tfe}.  In particular we will determine systematically
the NLO coefficients which relate the QCD operator definition of the
unpolarized and linearly polarized gluon TMD PDF with the unintegrated
gluon density of high energy factorization, which in turn will allow
us to recover the complete CSS resummation scheme in combination with
BFKL evolution, following closely related calculations based on
collinear factorization \cite{Ji:2005nu, Echevarria:2015uaa}. We expect our  result  to be useful for a precise
description of final states with small transverse momenta within high
energy factorized cross-sections, at a similar level of accuracy as
descriptions based on collinear factorization.
\\

Another aspect of our result relates to the derivation of TMD
splitting kernels in
\cite{Gituliar:2015agu,Hentschinski:2017ayz}. While the original
derivation was based on   a combined implementation of the
Curci-Furmanski-Petronzio formalism for the calculation of the
collinear splitting functions \cite{Curci:1980uw} and the framework of
high energy factorization provided by \cite{Lipatov:1995pn}, we find
in the following that the real contribution to the QCD operator
definition of the unpolarized gluon TMD yields precisely the
previously derived off-shell TMD splitting kernel. Our current study
provides therefore a possibility to recover the so far missing virtual
corrections to these off-shell splitting kernels.
\\

The outline of this paper is as follows. In Sec.~\ref{sec:setup} we
give a precise definition of the goal of this paper in more technical
terms, in particular the definition of the gluon TMD PDFs, Sec.~\ref{sec:eff} contains a brief review of Lipatov's high
energy effective action and presents among other details an extension
of the framework of \cite{Hentschinski:2020tbi} to
$k_T$-factorization. In Sec.~\ref{sec:gluonTMD} we present the results
of our NLO calculation, while Sec.~\ref{sec:evolution} discusses
aspects related to the interplay of CS and BFKL evolution. In
Sec.~\ref{sec:concl} we summarize our result and provide an outlook on
future research.
\\

\section{The setup of our study}
\label{sec:setup}

The starting point of our study is the TMD factorization of a suitable perturbative process. To be specific we will refer in the following to the transverse momentum distribution of a Higgs boson, as discussed for instance in \cite{Echevarria:2015uaa}, see also \cite{Ji:2005nu}. With $\pt_H$ and $m_H$ transverse momentum and mass of the Higgs boson, this factorization is valid for $|\pt_H| \ll m_H$ and reads:
\begin{widetext}
\begin{align}
  \label{eq:16Eche}
  \frac{d \sigma }{d y_H d^2 \pt_H} & =  \sigma_0(\mu) C_t^2(m_t^2, \mu) H(m_H^2, \mu) \int d^2 \qt_a d^2 \qt_b (2 \pi)^2 \delta^{(2)}(\pt_H - \qt_a - 
\qt_b)  \notag \\
& \hspace{6cm} 
2 \cdot x_A \Gamma^{ij}_{g/A}(x_A, \zeta_A; \qt_a, \mu) \cdot x_B \Gamma^{ij}_{g/B}(x_B, \zeta_B; \qt_b, \mu),
\end{align}
  \end{widetext}
where $y_H = 1/2 \ln (x_A/x_B)$ is the rapidity of the Higgs boson while $x_{A,B}$ denote the hadron momentum fractions of gluons stemming from hadron $A, B$ respectively and
\begin{align}
  \label{eq:zeta_scales}
  \zeta_{A,B} & = (p_H^\pm)^2 e^{\mp 2 y_c} = \left( M_H^2 + \pt_H^2\right)e^{\pm 2(y_H - y_c)},
\end{align}
where $y_c$ denotes the rapidity which divides soft gluons from hadron $A$ and $B$; $\mu$ is the renormalization point of the cross-section.  
To be specific we consider scattering of two hadrons with light-like momenta $p_A$ and $p_B$ which serve to define the light-cone directions 
\begin{align}
  \label{eq:nplusminus}
  (n^\pm)^\mu & = \frac{2}{\sqrt{{s}}} p_{A,B}^\mu, & {s} & = 2 p_A \cdot p_B ,
\end{align}
which yields the following Sudakov decomposition of a generic four-momentum, 
\begin{align}
k & = k^+ \frac{n^-}{2} +  k^- \frac{n^+}{2} + k_T, & k^\pm &= k \cdot n^\pm,
\end{align}
and $n^\pm \cdot k_T  = 0$. Here, $k_T$ is the embedding of the Euclidean vector $\kt$ into Minkowski space, so $k_T^2=-\kt^2$.
 For Eq.~\eqref{eq:16Eche}, the top quark is considered to be integrated out and $C_t$ is the corresponding Wilson coefficient;  $H$ is the square of the on-shell gluon form factor at time-like momentum transfer $q^2 = m_H^2$, with infrared divergences subtracted, see \cite{Ahrens:2008qu}. To leading order in perturbation theory they equal one, while the  precise NLO expression are not of interest for the following discussion and can be found for instance in \cite{Echevarria:2015uaa}. $\sigma_0(\mu)$ is finally the collinear Born level cross-section for the process $gg\to H$,
\begin{align}
  \label{eq:sigma0}
  \sigma_0 & = \frac{g_H^2 \pi}{8 (N_c^2-1)}, & g_H & = -\frac{\alpha_s(\mu)}{3 \pi v},
\end{align}
with $ v \simeq 246$~GeV the Higgs vacuum expectation value; $\alpha_s(\mu)$ denotes finally the strong coupling constant at the renormalization point $\mu$. For an unpolarized hadron,  the TMD correlator $\Gamma^{ij}$, $i,j = 1,2$  can be further decomposed \cite{Mulders:2000sh}
\begin{align}
  \label{eq:decompo4d}
  \Gamma^{ij}_{g/B}(x_B, \zeta_B; &\qt, \mu)  = - \frac{ \delta^{ij}}{2} f_{g/B}(x_B, \zeta_B; \qt_b, \mu) +
\notag \\ &
                                              \left(\frac{ \delta^{ij}}{2}  + \frac{\qt^i \qt^j}{\qt^2} \right)h_{g/B}(x_B, \zeta_B; \qt_b, \mu),
\end{align}
where $f_{g/B}(x_B, \zeta_B; \qt_b, \mu)$ denotes the unpolarized TMD gluon distriubtion and $h_{g/B}(x_B, \zeta_B; \qt_b, \mu)$ the linearly polarized TMD gluon distribution in an unpolarized hadron. 
In terms of QCD fields the TMD PDF is defined as
 \cite{Ji:2005nu, Echevarria:2015uaa}
 \begin{widetext}
 \begin{align}
  \label{eq:TMDdef}
    x \Gamma^{ij}_{g/B}(x_B, \zeta_B; \qt, \mu)  &=  
  \lim_{\sigma, y_n \to \infty } \int \frac{d \xi^+ d^2 {\bm \xi}}{2(2 \pi)^3 p_B^-} 
e^{i (x_B p_B^- \xi^+/2 - \qt \cdot \xit) } \, \tilde{\cal S}(2y_c, \sigma; \mu, \xit)
\notag \\
& \hspace{4cm}
 \cdot 
{\left \langle h(p_B) \left|\tr\left[\left(\mathcal{W}^{n(\sigma)}_\xi G^{-i}(\xi) \right)^\dagger \, \mathcal{W}^{n(\sigma)}_0 G^{-j}(0)\right] \right|h(p_B)\right \rangle} \bigg|_{\xi^- = 0} ,
\end{align}    
 \end{widetext}
 where $\tilde{\cal S}(2y_c, \sigma; \mu,\xit)$ denotes the soft factor and $\lim_{\sigma \to \infty}n(\sigma) = n^-$, with $\sigma \to \infty$ a suitable regulator whose  precise implementation  will be given in Eq.~\eqref{eq:n1n2} below.   Gauge links are in general given as a combination of a longitudinal  and a transverse gauge link \cite{Collins:2011zzd}, where the transverse gauge link is placed  at light-cone infinity. Working in covariant gauge, the gauge field at infinity vanishes and the transverse gauge link therefore equals  one. We will therefore in the following not consider the transverse gauge link. The longitudinal gauge link is on the other hand given by  
\begin{align}
  \label{eq:gauge_link}
  \mathcal{W}_\xi^n & = \mathrm{P} \exp \left(-\frac{g}{2} \int\limits_{- \infty}^{0}d \lambda n \cdot v(\lambda n   + \xi) \right),
 \end{align}
where $v_ \mu(x) = -i t^a v^a_\mu(x)$ denotes the gluonic field and
\begin{align}
  \label{eq:fieldstrength}
  D_\mu & = \partial_\mu + g v_ \mu,  & G^{\mu\nu} & = \frac{1}{g} \left[D^\mu, D^\nu \right]= -it^a G_{a}^{\mu\nu}.
\end{align}
 For the soft factor there exists various prescriptions in the literature, see {\it e.g.} \cite{Ji:2004wu, Ji:2005nu,Echevarria:2015uaa,Collins:2011zzd, Collins:2012uy,Echevarria:2012js}. To keep the discussion as general as possible, we will consider below the most general soft factor introduced in \cite{Collins:2011zzd, Collins:2012uy}, 
\begin{align}
  \label{eq:softC11}
 \tilde{ \mathcal{S}}(2y_c, \sigma; \mu, \xit) & = \sqrt{\frac{\tilde{S}(2y_c, 2y_n; \xit)}{\tilde{S}( \sigma, -2y_c,; \xit) \tilde{S}( \sigma, 2y_n; \xit)}},
\end{align}
with
\begin{align}
  \label{eq:Sy1y2_def}
 & \tilde{S}(y_1, y_2; \xit)  = \frac{1}{N_c^2 -1} \notag \\
  &\cdot \left \langle 0\left| (\mathcal{W}_\xit^{n_1(y_1)})^\dagger \mathcal{W}_\xit^{n_2(y_2)}(\mathcal{W}_\xit^{n_2(y_2)})^\dagger \mathcal{W}_\xit^{n_1(y_1)}\right|0\right\rangle,
\end{align}
where $ n_{1,2}(y_{1,2})$ are tilted Wilson lines such that $n_1$ is
placed at rapidity\footnote{Due to the regulator defined in
  Sec.~\ref{sec:regularization} this implies in our case also an
  imaginary part; the above expressions refer to the corresponding
  real parts, while imaginary parts cancel for the final result}
$y_1/2$ and $n_2$ at rapidity $-y_2/2$. For a precise definition of
the light-cone directions see Eq.~\eqref{eq:n1n2} further below.  The
goal of the following sections is to study this gluon TMD in the high
energy limit at next-to-leading order. In particular we will discuss
the factorization of this TMD PDF into a perturbative coefficient, the
BFKL gluon Green's function, and a hadronic impact factor. The latter
two will then form the so-called unintegrated gluon density within
high energy factorization. Our study is limited to the exchange of 2
reggeized gluons. It is known from various studies, that the gluon TMD
also receives corrections due to the exchange of multiple reggeized
gluons, which are of importance to take into account corrections due
to high gluon densities and their possible saturation, see {\it e.g.}
\cite{Dominguez:2011wm,Cali:2021tsh,Altinoluk:2021ygv,
  vanHameren:2020rqt, Stasto:2018rci}. While these are very
interesting questions -- in particular since they can provide
modifications of the region of very small transverse momenta due to
the emergence of a saturation scale -- we do not consider these
effects in the following. Instead, great care will be taken to provide
a complete discussion of various factorization scales and parameters
both due to factorization in the soft-limit and the high energy limit,
as well as UV renormalization. The study of these effects is somehow
more straightforward, if the observable is restricted to 2 reggeized gluon
exchange, which is the reason why we focus on this limit in the
following. The obtained results may then be generalized at a later
stage to the case of multiple reggeized gluon exchange.
\\

Another motivation for this study is to link the above gluon TMD to the TMD splitting kernels derived in \cite{Gituliar:2015agu,Hentschinski:2017ayz}. Below we will demonstrate that the TMD gluon-to-gluon splitting of \cite{Hentschinski:2017ayz} arises directly from the real contributions to the 1-loop coefficient. We believe that this is a very interesting result, since it allows us to connect the framework of real TMD splitting kernels to the above operator definitions of TMD PDFs.

\section{The High-Energy Effective Action}
\label{sec:eff}

\begin{figure*}[t]
    \label{fig:subfigures}
  % \centering
\includegraphics[width=\textwidth]{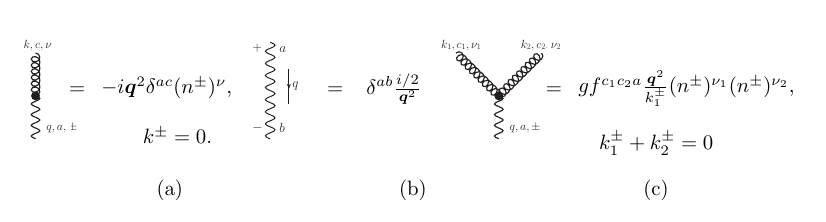}
 \caption{ Feynman rules for the lowest-order effective vertices of the effective action. Wavy lines denote reggeized fields and curly lines gluons. }
\label{fig:3}
\end{figure*}
Since the current study requires a small but important generalization in comparison to the framework presented in \cite{vanHameren:2020rqt}, we begin our study with a short review of the high energy effective action and the resulting calculational framework for NLO calculations. Our treatment of high energy factorization is based on  Lipatov's high energy effective action
\cite{Lipatov:1995pn}. Within this framework, QCD amplitudes are in
the high energy limit decomposed into gauge invariant sub-amplitudes
which are localized in rapidity space and describe the coupling of
quarks ($\psi$), gluon ($v_\mu$) and ghost ($\phi$) fields to a new
degree of freedom, the reggeized gluon field $A_\pm (x)$. The latter
is introduced as a convenient tool to reconstruct the complete QCD
amplitudes in the high energy limit out of the sub-amplitudes
restricted to small rapidity intervals. 
%%%%%%%%%%%%%%%%%%%%%%%%%%%%%%%%%%%%%%%%
Lipatov's effective action is then obtained by adding an induced term $
S_{\text{ind.}}$ to the QCD action $S_{\text{QCD}}$,
\begin{align}
  \label{eq:effac}
S_{\text{eff}}& = S_{\text{QCD}} +
S_{\text{ind.}}\; ,
\end{align}
where the induced term $ S_{\text{ind.}}$ describes the coupling of
the gluonic field $v_\mu = -it^a v_\mu^a(x)$ to the reggeized gluon
field $A_\pm(x) = - i t^a A_\pm^a (x)$.  High energy factorized
amplitudes reveal strong ordering in plus and minus components of
momenta which is reflected in the following kinematic constraint
obeyed by the reggeized gluon field:
\begin{align}
  \label{eq:kinematic}
  \partial_+ A_- (x)& = 0 = \partial_- A_+(x).
\end{align}
Even though the reggeized gluon field is charged under the QCD gauge
group SU$(N_c)$, it is invariant under local gauge transformation
$\delta A_\pm = 0$.  Its kinetic term and the gauge invariant coupling
to the QCD gluon field are contained in the induced term
\begin{align}
\label{eq:1efflagrangian}
  S_{\text{ind.}} = \int &\text{d}^4 x \,
\text{tr}\left[\left(W_-[v(x)] - A_-(x) \right)\partial^2_\perp A_+(x)\right]\notag \\
&+\text{tr}\left[\left(W_+[v(x)] - A_+(x) \right)\partial^2_\perp A_-(x)\right],
\end{align}
with 
\begin{align}
  \label{eq:funct_expand}
  W_\pm[v(x)] =&
v_\pm(x) \frac{1}{ D_\pm}\partial_\pm,% =  
%- \frac{1}{g} \partial_\pm  U[v_\pm]
%= v_\pm - g  v_\pm\frac{1}{\partial_\pm} v_\pm + g^2 v_\pm
%\frac{1}{\partial_\pm} v_\pm\frac{1}{\partial_\pm} v_\pm - \ldots
&
D_\pm & = \partial_\pm + g v_\pm (x).
\end{align}
For a more in depth discussion of the effective action we refer to the
 reviews \cite{Chachamis:2012mw, Hentschinski:2020rfx}. Due to the induced term in
Eq.~(\ref{eq:effac}), the Feynman rules of the effective action
comprise, apart from the usual QCD Feynman rules, the propagator of
the reggeized gluon and an infinite number of so-called induced
vertices.  Vertices and propagators needed for the current study are
collected in Fig.~\ref{fig:3}.
Determination of NLO corrections using this effective action approach
has been addressed recently in series of publications
\cite{Hentschinski:2011tz, Chachamis:2012cc,
  Chachamis:2012gh,Chachamis:2013hma, Hentschinski:2011xg}. For a
discussion of the analogous high energy effective for flavor exchange
\cite{Lipatov:2000se} at NLO see {\it e.g.}
\cite{Nefedov:2017qzc,Nefedov:2018vyt,Nefedov:2019mrg}.

\subsection{Determination of NLO coefficients}
\label{sec:frameworkNLO}

\begin{figure*}[t]
  \centering
  \parbox{.4\textwidth}{\includegraphics[width=0.3\textwidth]{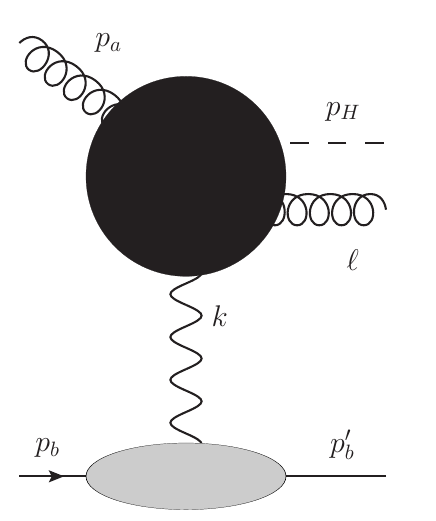}}
  \parbox{.4\textwidth}{\includegraphics[width=0.3\textwidth]{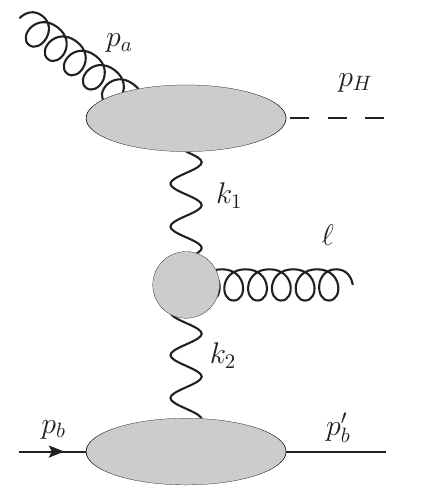}}
  \caption{NLO correction to forward Higgs production (left) and the factorized matrix element with internal reggeized gluon exchange (right). The NLO correction contains the factorized contribution, which need to be subtracted from the former.}
  \label{fig:Higgs_schem}
\end{figure*}
The framework for the determination of NLO corrections has been established in \cite{Hentschinski:2020tbi} within the determination of the NLO forward Higgs production coefficient in the infinite top mass limit.  We will therefore frequently refer to  process
\begin{align}
  \label{eq:example_process}
  \text{gluon}(p_a) + \text{quark}(p_b) \to \text{Higgs}(p_H) + X,
\end{align}
as an example process in the following, where we further assume that
the particles in the fragmentation region of the scattering particles
are widely separated in rapidity. The partonic impact factor of the
quark with momentum $p_b$ will be later on replaced by the hadronic
impact factor, which forms together with the BFKL Green's function the
unintegrated gluon density. As in \cite{Hentschinski:2020tbi}, we will
consider matrix elements  normalized to match
corresponding collinear matrix elements for vanishing virtuality of the reggeized gluon state. We therefore have
\begin{align}
  \label{eq:dsig_hat}
 &  \overline{|\mathcal{A}_{ar^+  \to   X_a^{(n)}}|^2}
=
 \frac{(k^-)^2}{4 \kt^2}\overline{|\mathcal{M}_{ar^+  \to   X_a^{(n)}}|^2}; 
\end{align}
where we average over incoming parton color as well as the color of the reggeized gluon and sum over the color of produced particles; $X_{a}^{(n)}$ denotes the $n$-particle system produced in the regarding fragmentation region. With
\begin{align}
  \label{eq:phasepsace}
d\Gamma^{(n)} & =   (2 \pi)^d \delta^d\left(p_a + k - \sum_{j=1}^n p_j\right) d \Phi^{(n)}, \notag \\
 d \Phi^{(n)} & = \prod_{j=1}^n \frac{d^d p_j}{(2 \pi)^{d-1}}  \delta_+\left(p_j^2 - m_j^2\right),
\end{align}
we arrive at the following definition of an off-shell partonic cross-section $ d\hat{\sigma}_{a+}$ and the corresponding impact factor $\hat{h}^{k_T}(\kt)$:
\begin{align}
  \label{eq:sigma}
 d\hat{\sigma}_{a+} & =  \frac{ \overline{|\mathcal{A}_{ar^+  \to   X_a^{(n)}}|}^2}{2 p_a^+ k^-} d\Gamma^{(n)},
\notag \\
\hat{h}^{k_T}(\kt) & =    \int \frac{dk^-}{k^-} d\hat{\sigma}_{a+} .
\end{align}
Note that this impact factor can  in principle  be arbitrarily differential, as far as the formulation of high energy factorization is concerned; for a corresponding definition of the other impact factor we refer to  \cite{Hentschinski:2020tbi}. The above expression is subject to so-called rapidity divergences
which  are 
understood to be regulated through lower cut-offs on the rapidity of
all particles, $\eta_i > -\rho/2$ with $\rho \to \infty$ and
$i = 1, \ldots, n$ for $n$ the number of particles produced in the
fragmentation region of the initial parton $a$. For virtual
corrections, the regularization is achieved  through tilting
light-cone directions of the high energy effective action,
\begin{align}
  \label{eq:tilting}
  n^+ & \to n^+ + e^{- \rho} n^-, & \rho & \to \infty. 
\end{align}
Below we will also comment on the possibility to regularize rapidity
divergences through tilting light-cone direction also in the case of
real corrections, see Sec.~\ref{sec:regularization}.
\\

As shown through explicit results \cite{Hentschinski:2011tz,
  Chachamis:2012cc, Chachamis:2012gh,Chachamis:2013hma,
  Hentschinski:2020tbi}, impact factors contain beyond leading order
configurations, which reproduce factorized contributions with internal
reggeized gluon exchange. It is therefore necessary to subtract these
contributions, see also Fig.~\ref{fig:Higgs_schem}. To this end one
defines the bare one-loop 2-reggeized-gluon Green's function
$G_B(\kt_1,\kt_2)$ as well as the impact factors through the following
perturbative expansion
\begin{align}
  \label{eq:Uuu}
  G_B(\kt_1, \kt_2; \rho)
& =  \delta^{(2 + 2 \epsilon)} (\kt_1 + \kt_2) +  G_B^{(1)}(\kt_1, \kt_2; \rho) + \ldots
 \notag \\
  h_{a}(\kt, \rho) & =  h_{a}^{(0)}(\kt) +  h_{a}^{(1)}(\kt, \rho) + \ldots .
%G_B^{(1)}(\kt_1, \kt_2; \rho)  & = \frac{\alpha_s C_A \rho}{\pi} \left( \frac{1}{\pi_ \epsilon (\kt_1 - \kt_2)^2} - \frac{1}{\epsilon} \left(\frac{\kt_1^2}{\mu^2} \right)^\epsilon\delta^{(2 + 2 \epsilon)} (\kt_1 - \kt_2) \right)\\
%& -
%\frac{\alpha_s}{2 \pi}  \left(\frac{\kt_1^2}{\mu^2} \right)^\epsilon \left(\frac{5 C_A - 2 n_f}{3 \epsilon} - \frac{31 C_A}{9} + \frac{10 n_f}{9} \right)   \delta^{(2 + 2 \epsilon)} (\kt_1 - \kt_2),
\end{align}
%where
%\begin{align}
%  \label{eq:somedef}
%  \pi_\epsilon& \equiv \pi^{1 + \epsilon} \Gamma(1 - \epsilon)\mu^{2 \epsilon}, & \alpha_s & = \frac{g^2 \Gamma(1 - \epsilon) \mu^{2 \epsilon}}{(4 \pi)^{1 + \epsilon}}.
%\end{align}
%Note that the rapidity divergence is directly proportional to the leading order BFKL kernel.
Using the following convolution convention
\begin{align}
  \label{eq:64}
\left[  f \otimes g\right](\kt_1, \kt_2) & \equiv  \int d^{2 + 2 \epsilon}  \qt 
f(\kt_1, \qt) g(\qt, \kt_2),
\end{align}
we then   define  the following subtracted bare NLO coefficient,
\begin{align}
  \label{eq:coeff}
   C_{a,B}^{(1)}(\kt, \rho)  & =   h_{a}^{(0)}(\kt) +  h_{a}^{(1)}(\kt, \rho)
                               \notag \\
  & \qquad -
 \left[h_{a}^{(0)} \otimes G^{(1)}_B(\rho) \right](\kt),
\end{align}
and \footnote{Note that the impact factors themselves might depend on
  additional transverse momenta; this is however irrelevant for the
  following discussion of high energy factorization and we therefore
  suppress this dependence in the following.}
\begin{align}
  \label{eq:dsigNLO}
  d \sigma_{ab}^{\text{NLO}} & = \frac{1}{\pi^{1+ \epsilon}}\left[ C^{k_T}_{a,B}(\rho) \otimes G_B(\rho) \otimes C^{(ugd)}_{b,B}(\rho) \right] \notag \\
  & \qquad + \text{terms beyond NLO},
\end{align}
where we added the super-scripts `$k_T$' and `$ugd$' to indicate that the impact factor of the particle with momentum $p_{a,b}$ refers to the hard event and the unintegrated gluon distribution respectively. 
While rapidity divergences cancel for the the above expression,   its elements still depend on the regulator.  As a next step we therefore define  a renormalized Green's function $G_R$  through
\begin{align}
  \label{eq:58}
  G_B(\kt_1,& \kt_2; \rho)  =  \bigg[  Z^+\left(\frac{\rho}{2} -\eta_a\right) \notag \\
  & \otimes G_R\left( \eta_a - \eta_b\right)   \otimes Z^- \left(\frac{\rho}{2}+ \eta_b\right) \bigg] (\kt_1, \kt_2),
\end{align}
which yields
\begin{align}
  \label{eq:dsig_renom}
  d\sigma_{ab}^{\text{NLO}} = \left[C_{a,R}(\eta_a) \otimes G_R(\eta_a, \eta_b) \otimes C_{b,R}(\eta_b) \right] ,
\end{align}
where
\begin{align}
  \label{eq:CRs}
  C_{a,R}(\eta_a;\kt_1) & \equiv \left[ C_a(\rho)  \otimes Z^+\left(\frac{\rho}{2} - \eta_a\right)\right](\kt_1), \notag \\
 C_{b,R}(\eta_b;\kt_2) & \equiv \left[  Z^+\left(\frac{\rho}{2} +\eta_b\right) \otimes C_b (\rho) \right](\kt_2)  \, .
\end{align}
The transition functions $Z^\pm$ have a twofold purpose: they both serve to cancel $\rho$-dependent terms between impact factors and Green's function and allow to define the BFKL kernel.  In particular,
\begin{align}
  \label{eq:BFKL_forZ}
  \frac{d}{d\hat{\rho}} Z^+(\hat{\rho}; \kt, \qt)  & =  
\left[  Z^+(\hat{\rho}) \otimes K_{\text{BFKL}} \right] (\kt, \qt), \notag \\
\frac{d}{d\hat{\rho}} Z^-(\hat{\rho}; \kt, \qt)  & =  
\left[ K_{\text{BFKL}} \otimes  Z^-(\hat{\rho})\right] (\kt, \qt),
\end{align}
where
\begin{align}
  \label{eq:62}
  K_{\text{BFKL}}(\kt, \qt) & =  K^{(1)}(\kt, \qt) +  K^{(2)}(\kt, \qt) + \ldots \,.
\end{align}

\subsection{Transition function and finite terms}
\label{sec:finite_trans}

In the following we generalize the treatment given in
\cite{Hentschinski:2020tbi} through including as well the most general
finite contribution into our discussion. The need to include finite
contribution into this transition factors has been first realized in
the determination of the 2-loop gluon Regge trajectory in
\cite{Chachamis:2013hma,Chachamis:2012gh,Chachamis:2012cc}, where both
divergent and finite terms could be simply taken to exponentiate,
since one is dealing with 
1-reggeized gluon exchange contributions only. A suitable
generalization, which both reduces to the exponential ansatz of
\cite{Chachamis:2013hma} and obeys Eq.~\eqref{eq:BFKL_forZ}, is then
given by
\begin{align}
  \label{eq:61gen}
    Z^+&(\hat{\rho}; \kt, \qt)   = 
             \delta^{(2 + 2 \epsilon)}(\kt - \qt) + \hat{\rho}
                                 K_{\text{BFKL}}(\kt, \qt)
                                 \notag \\
  &
                                 + 
\frac{\hat{\rho}^2}{2} K_{\text{BFKL}}\otimes  K_{\text{BFKL}}  (\kt, \qt)
 + 
f^+ \otimes \hat{\rho} K_{\text{BFKL}}  (\kt, \qt)
\notag \\ & + 
f^+(\kt, \qt) + \frac{ f^+ \otimes f^+(\kt, \qt)}{2} \ldots , \notag \\
 Z^-&(\hat{\rho}; \kt, \qt) 
  = 
             \delta^{(2 + 2 \epsilon)}(\kt - \qt) + \hat{\rho}
      K_{\text{BFKL}}(\kt, \qt)
      \notag \\
  &
      + 
\frac{\hat{\rho}^2}{2} K_{\text{BFKL}}\otimes  K_{\text{BFKL}}  (\kt, \qt)
% \notag \\
% &
    + 
 \hat{\rho} K_{\text{BFKL}} \otimes f^- (\kt, \qt)
    \notag \\
  &
    + 
f^-(\kt, \qt) + \frac{ f^- \otimes f^-(\kt, \qt)}{2} \ldots ,
\end{align}  
which is sufficient for a discussion up to NLO accuracy. As we will see in the following, these finite contributions serve a twofold purpose. At first they  remove a potential finite contribution in the bare Green's function, 
\begin{widetext}
\begin{align}
\label{eq:1}
G_B^{(1)}(\kt_1, \kt_2; \rho)  & = \frac{\alpha_s C_A \rho}{\pi} \left( \frac{1}{\pi_ \epsilon (\kt_1 - \kt_2)^2} - \frac{1}{\epsilon} \left(\frac{\kt_1^2}{\mu^2} \right)^\epsilon\delta^{(2 + 2 \epsilon)} (\kt_1 - \kt_2) \right) \notag \\
& \hspace{1.5cm}-
\frac{\alpha_s}{2 \pi}  \left(\frac{\kt_1^2}{\mu^2} \right)^\epsilon \left(\frac{5 C_A - 2 n_f}{3 \epsilon} - \frac{31 C_A}{9} + \frac{10 n_f}{9} \right)   \delta^{(2 + 2 \epsilon)} (\kt_1 - \kt_2),
\end{align}
  \end{widetext}
where
\begin{align}
  \label{eq:somedef}
  \pi_\epsilon& \equiv \pi^{1 + \epsilon} \Gamma(1 - \epsilon)\mu^{2 \epsilon}, & \alpha_s & = \frac{g^2 \Gamma(1 - \epsilon) \mu^{2 \epsilon}}{(4 \pi)^{1 + \epsilon}},
\end{align}
which then yields directly the 1-loop BFKL kernel
\begin{align}
  \label{eq:BFKL1}
  K^{(1)}(\kt_1, \kt_2)  &= \frac{\alpha_s C_A }{\pi} \bigg[ \frac{1}{\pi_ \epsilon (\kt_1 - \kt_2)^2} \notag \\
  & \qquad - \frac{1}{\epsilon} \left(\frac{\kt_1^2}{\mu^2} \right)^\epsilon\delta^{(2 + 2 \epsilon)} (\kt_1 - \kt_2) \bigg].
\end{align}
To keep the treatment as general as possible, the finite contribution is on the other hand then split up into 2 terms:
\begin{align}
  \label{eq:finite_spli}
    f^{\pm,(1)}\left( \kt_1, \kt_2 \right) & =  \tilde{f}^{\pm,(1)}\left( \kt_1, \kt_2 \right)+  \bar{f}^{\pm,(1)}\left( \kt_1, \kt_2 \right),
\end{align}
The contribution $\tilde{f}^\pm$ is used to transfer finite terms contained in the Green's function to the impact factors. We take in the following this function to be identical for both plus and minus direction $\tilde{f}^{\pm,(1)}\left( \kt_1, \kt_2 \right)= \tilde{f}^{(1)}\left( \kt_1, \kt_2 \right)$. Indeed, since  we are  essentially dealing with contributions due to the gluon polarization tensor in the high energy limit, such a symmetric treatment appears to be the appropriate one. One finds
\begin{align}
  \label{eq:2}
\tilde{f}^{(1)}&\left( \kt_1, \kt_2 \right)  =  \frac{\alpha_s}{4 \pi} \bigg[
 - \frac{5 C_A - 2n_f}{3\epsilon} \left(\frac{\kt_1^2}{\mu^2}\right)^\epsilon \notag \\
& +
 \frac{31 C_A - 10 n_f}{9}  \bigg] \delta^{(2 + 2 \epsilon)}(\kt_1 - \kt_2) + \mathcal{O}(\epsilon).
\end{align}
The second contribution $\bar{f}^\pm$ was absent in the discussion of \cite{Hentschinski:2020tbi}. It needs to satisfy to  1-loop the following requirement, to ensure absence of finite terms in the Green's function:
\begin{align}
  \label{eq:3}
  \bar{f}^{+,(1)}\left( \kt_1, \kt_2 \right) & = - \bar{f}^{-,(1)}\left( \kt_1, \kt_2 \right) \equiv   \bar{f}^{(1)}\left( \kt_1, \kt_2 \right).
\end{align}
Note that similar constraints can be imposed on higher order contributions to the functions $\bar{f}^\pm$ to achieve a Green's function without finite contributions at higher orders. Including all contributions, one finally arrives at  following expression for the NLO coefficient:
\begin{align}
  \label{eq:4}
  {C}_{R, i}^{\text{NLO}}(\kt) &= %C_R^{\text{NLO}}(\kt) + h_a^ 0 \otimes {f}^{+, (1)} (\kt) \notag \\
  h_a^{(0)}(\kt) + h_a^{(1)}(\kt) + h_a^{(0)} \notag \\
  & \otimes \left[(-\frac{\rho}{2}- s_i\eta_i) K^{(1)}+ s_i  \bar{f}^{ (1)} - \tilde{f}^{(1)} \right](\kt)\notag \\
i & = a,b, \qquad s_{a,b} = \pm,
\end{align}
where $K^{(1)}(\kt_1, \kt_2)$ and $\tilde{f}^{(1)}(\kt_1, \kt_2)$ are
given in Eq.~\eqref{eq:BFKL1} and Eq.~\eqref{eq:2} respectively. The
function $\bar{f}^{(1)}(\kt_1, \kt_2)$ as well as the evolution
parameters $\eta_{a,b}$ are on the other hand still undetermined. They
are in principle arbitrary, but should be chosen such that both impact
factors are free of large logarithms.

\subsection{Scale setting for $k_T$-factorization}
\label{sec:kT_scale}

The parameters $\eta_{a,b}$ as well as the function $\bar{f}$ are at
first arbitrary; the former define through the combination
$\eta_a - \eta_b$ the evolution parameter of the 2 reggeized gluon
Green's function $G_R$, see also the related discussion in
\cite{Hentschinski:2020tbi, Hentschinski:2020rfx}.  As usually, it is
necessary to chose these parameters such that the next-to-leading
order corrections to impact factors are under perturbative control and
that large logarithms in the center of mass energy are resumed through
the 2 reggeized gluon Green's function. Within the $k_T$-factorization
setup, one of the impact factors, {\it e.g.} the coefficient
$C_{R,b}(\eta_b, \kt_2)$ in our example, is to be replaced by the
hadronic impact factor, which then builds together with the BFKL
Green's function the unintegrated gluon density. Even though the
hadronic impact factor is naturally a non-perturbative object, at the
very least for transverse momenta $|\kt_2| \gtrsim 1.5$~GeV, it must
have an expansion in terms of collinear parton distribution functions
and corresponding collinear coefficients, see {\it e.g.}
\cite{Ciafaloni:1998hu}.  In order to have a complete matching of the
resulting expression for the unintegrated gluon density to the
collinear gluon distribution in the double-logarithmic limit, it is
natural to chose the evolution parameter $\eta_a - \eta_b$ to coincide
with the fraction of the hadronic momentum carried on by the gluon $x \simeq M_a^2/s$ with $M_a$ the invariant mass of the system produced in the fragmentation region\footnote{This relation is exact for the scattering of a parton $a$ with a hadron $B$}. Note that at NLO, an asymmetric scale choice, {\it i.e.} choosing the
reference scale of the center-of-mass energy $\sqrt{s}$ to be of the
order of a typical scale of \emph{one} of the impact factors, leads to
a modification of the NLO BFKL kernel, see {\it e.g.}
\cite{Fadin:1998py,Bartels:2006hg}. To repeat this exercise within the
context of the high energy effective action, we reconsider
Eq.~\eqref{eq:CRs}, but focuse now on the factorization parameters
$\eta_{a,b}$ and the finite terms introduced above:
\begin{align}
  \label{eq:8}
  {C}_{R,a}^{\text{NLO}}&\left(\kt_1, M_a, \eta_a, \bar{f}^{(1)}\right) = 
                                                                         \hat{C}_{R,a}^{\text{NLO}}(\kt_1, M_a) \notag \\
  & -\left[h_a^{(0)}(M_a) \otimes \left(\eta_a K^{(1)} - \bar{f}^{(1)} \right) \right](\kt_1) \notag \\
 {C}_{R,b}^{\text{NLO}}&\left(\kt_2, M_b, \eta_b,\bar{f}^{(1)}\right) = 
\hat{C}_{R,b}^{\text{NLO}}(\kt_2, M_a) \notag \\ &+\left[h_b^{(0)}(M_a) \otimes \left(\eta_b K^{(1)} - \bar{f}^{(1)} \right) \right](\kt_2)
%\notag \\ \bar{C}_R^{+,\text{NLO}}(\kt_1, M_a, \eta_a) &= 
%\hat{C}_R^{\text{NLO}}(\kt_1, M_a) \notag \\
%& \hspace{1cm}- \int d^{2 + 2 \epsilon} \qt_1  h_a^{(0)}( \qt_1, M_1)  \left[\eta_a K^{(1)}(\qt_1, \kt_1) -  \bar{f}^{ (1)} (\qt_1, \kt_1)  \right]
% \notag \\
%\bar{C}_R^{-,\text{NLO}}(\kt_2, M_b, \eta_b) &= \hat{C}_R^{\text{NLO}}(\kt_2, M_b) 
%\notag \\
%& \hspace{1cm} + \int d^{2 + 2 \epsilon} \qt_2  \left[\eta_b K^{(1)}(\kt_2, \qt_2) -  \bar{f}^{ (1)} (\kt_2, \qt_2)  \right]  h_a^{(0)}( \qt_2, M_2);
\end{align}
where $M_{a,b}$ is the invariant mass of the produced final state corresponding to each of  the impact factors and  $\hat{C}_R$ collects all terms which are independent of both  $\eta_{a,b}$ and $ \bar{f}$. To equal the evolution parameter of the Green's function with the hadron momentum fraction carried on by the reggeized gluon entering the impact factor ${C}_{R,a}$, we set
\begin{align}
  \label{eq:9}
  \eta_a& = \ln \frac{M_0}{k^-}, \qquad  \eta_b = \ln \frac{M}{p_b^-}, \notag \\  \eta_a - \eta_b & = \ln \frac{p_b^-}{k^-} + \ln \frac{M_0}{M} = \ln \frac{x_0}{x_g},
\end{align}
where $M, M_0$ are so far unspecified reference scale and
$x_0 = M_0/M$ is a parameter of order one, which allows to estimate
the scale uncertainty associated with high energy factorization. In
the following we chose $M$ to be of the order of $M_a$, {\it i.e.} the
hard scale. While this is a natural choice for the hard impact factor,
it introduces the same scale into the hadronic impact factor,
characterized in general by small transverse momenta. We therefore find in the perturbative region of the hadronic impact factor a  large collinear logarithm, which at first spoils the
convergence of the perturbative expansion.  This logarithm can however be 
absorbed into the $\bar{f}$ function through setting
\begin{align}
  \label{eq:11}
  \bar{f}_{k_T}^{(1)}(\kt_2, \qt_2) & \equiv  \ln \frac{M}{|\kt_2|} K^{(1)}(\kt_2, \qt_2),
\end{align}
which eliminates the logarithm in $M$ from the hadronic impact factor. Note that the choice of $|\kt_2|$ as the relative scale is somewhat arbitary, and other choices are equally possible, see {\it e.g.} \cite{Ciafaloni:1998hu}.
It is interesting to compare this situation to the case where the parameters $\eta_{a,b}$ are identified with the rapidities of the system produced in the regarding fragmentation region, $\eta_{a} = \ln p_a^+/M_a$,  $\eta_{b} = \ln M_b/p_b^-$, with $\bar{f}^{(1)}=0$. One finds
\begin{align}
  \label{eq:8YY}
   {C}_{R,b}^{\text{NLO}}&\left(\kt_2, M_b, \ln \frac{M}{p_b^-}, \bar{f}_{k_T}^{(1)} \right) = {C}_{R,b}^{\text{NLO}}\left(\kt_2, M_b, \ln \frac{M_b}{p_b^-}, 0 \right) \notag \\
&-  \ln \frac{M_b}{|\kt_2|} \int d^{2 + 2 \epsilon} \qt_2 \left[  K^{(1)}(\kt_2, \qt_2)    h_b^{(0)}( M_b, \qt_2) \right],
\notag \\
  {C}_{R,a}^{\text{NLO}}&\left(\kt_1, M_a, \ln \frac{M_0}{k^-}, \bar{f}_{k_T}^{(1)} \right) = {C}_{R,a}^{\text{NLO}}\left(\kt_1, M_a, \ln \frac{p_a^+}{M_a}, 0 \right) \notag \\
                                                                                          & \hspace{-.5cm}+  \int d^{2 + 2 \epsilon} \qt_1 \left[h_a^{(0)}( M_a, \qt_a) \cdot    \ln \frac{ M_a}{x_0 |\qt_1|} \cdot  K^{(1)}(\qt_1, \kt_1)     \right],
\end{align}
which allows to verify that the presented treatment agrees -- after setting $x_0 = 1$ -- with the one derived  in  \cite{Bartels:2006hg}, based on a study of ladder diagrams within the Quasi-Multi-Regge-Kinematics in the context of the definition of the NLO inclusive jet vertex. For  the NLO BFKL kernel one finally obtains the following contribution
\begin{align}
  \label{eq:12}
  {K}_{k_T}^{(2)}&(\kt_1, \kt_2)  =  {K}^{(2)}(\kt_1, \kt_2)  \notag \\
  &- \frac{1}{2}  \int d^{2} \kt \ln \frac{\kt^2}{\kt_1^2} K^{(1)}(\kt_1, \kt) K^{(1)}(\kt, \kt_2),
\end{align}
which is independent of the parameter $x_0$ and where $ {K}^{(2)}$ denotes the NLO BFKL kernel if $\eta_{a,b}$ is identified with the rapidities of the external particles with $\bar{f}^{(1)}=0$. The above expression is in agreement with   \cite{Bartels:2006hg} and  \cite{Fadin:1998py}.   A more  detailed discussion of possible choices of the function $\bar{f}^{(1)}$ will be presented elsewhere.  \\

Summing up we have the general definition of the unintegrated gluon density
\begin{align}
  \label{eq:ugd_general}
  \mathcal{G}&\left(\Delta \eta_{ab}, \eta_b, \kt, \bar{f}^{(1)}\right) \notag \\ & = \int d^{2 + 2 \epsilon} \qt G_R(\Delta \eta_{ab}, \kt, \qt) {C}^{\text{NLO}}_{R,b} \left(\qt, \eta_b, \bar{f}^{(1)} \right),
\end{align}
where we suppressed the dependence on the invariant mass $M_b$ since the latter can in general be expressed in terms of the transverse momentum. The $k_T$-factorization scheme fixes then $\bar{f}^{(1)}$ through Eq.~\eqref{eq:11}, while $\Delta \eta_{ab} = \eta_a - \eta_b$ is set to $\Delta \eta^{k_T}_{ab} \equiv \ln1/x_g$. The high energy factorized cross-section is then obtained as
\begin{align}
  \label{eq:dsig}
  d\sigma_{AB} & =  \int \frac{d^{2 + 2 \epsilon} \kt}{\pi^{1+ \epsilon}}   d{C}_{R,a}^{\text{NLO}}\left(\kt, M_a, \eta_a, \bar{f}^{(1)}\right) \notag \\
  & \hspace{3cm} \mathcal{G}\left(\Delta \eta_{ab}, \eta_b, \kt, \bar{f}^{(1)}\right),
\end{align}
where `$A$'might either denote a parton $a$, a partonic impact factor convoluted with a parton distribution function of a hadron $A$ or a colorless initial state which allows for a perturbative treatment. Concluding we remark that in \cite{Altinoluk:2019wyu} a definition of the unintegrated gluon density has been proposed in terms of a operator definition with the high energy gluonic field in light-cone gauge, which requires the inclusion of both so-called 2,3, and 4 body contributions. While an interpretation of such contributions in terms of induced vertices Fig.~\ref{fig:3} of the high energy effective action appears to be possible, the precise relation remains unclear.

\subsection{Regularization of rapidity divergences}
\label{sec:regularization}

\begin{figure*}[t]
  \centering
  \parbox{.21\textwidth}{\includegraphics[width=.21\textwidth]{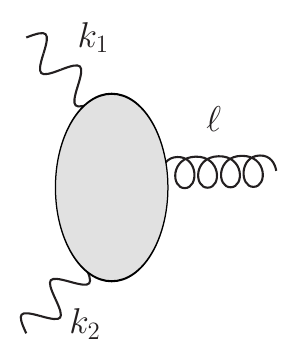}}
   \parbox{.16\textwidth}{ $\,$}
 \parbox{.2\textwidth}{\includegraphics[width=.2\textwidth]{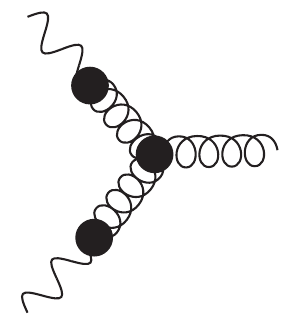}}
 \parbox{.2\textwidth}{\includegraphics[width=.2\textwidth]{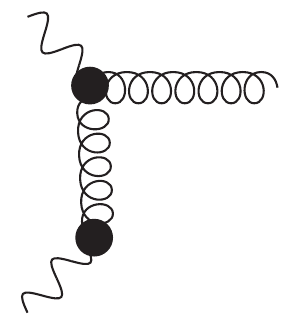}}
 \parbox{.2\textwidth}{\includegraphics[width=.2\textwidth]{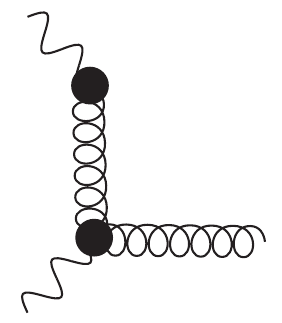}}
 
  \caption{Real emission contribution to the Lipatov vertex which yields the  1-loop Green's function (left) and contributing Feynman diagrams (right). Note current conservation of the Lipatov vertex is given both for tilted and untilted light-cone directions.}
  \label{fig:central}
\end{figure*}

As pointed out in the above discussion, high energy factorized matrix
elements are subject to so-called rapidity divergences. While they
cancel at the level of observables after subtraction of factorizing
contribution and use of the transition function, intermediate results
beyond leading order require a regulator in order to arrive at
well-defined matrix elements. While for real production a cut-off on
the rapidity of produced particles provides a natural way to regulate
such divergences, a consistent regularization is more difficult to
achieve in the case of virtual diagrams. As already pointed out in
Sec.~\ref{sec:frameworkNLO}, a suitable way to regulate these
divergences in the case of virtual corrections is to tilt the
light-cone directions of the high energy effective action away from
the light-cone, see Eq.~\eqref{eq:tilting}.  Note that from a formal point of view this a very attractive way of
regularizing rapidity divergences, since gauge invariance of the high
energy effective action does not depend on the property
${n^\pm}^2 = 0$; tilting light cone directions provides therefore a
gauge invariant regulator, similar to dimensional
regularization. Nevertheless the current treatment, see {\it e.g.}
\cite{Hentschinski:2011tz,Chachamis:2012cc,Hentschinski:2020tbi} is
somewhat unsatisfactory, since it treats real (cut-off) and virtual
(tilting) corrections on somewhat different grounds. At the same time,
tilting light-cone directions is also a frequently used regulator for
the determination of the 1-loop corrections to TMD PDFs within
collinear factorization, see {\it e.g.} \cite{Ji:2005nu,
  Collins:2011zzd} and references therein, which is being use for both
real and virtual corrections. It seems therefore natural to regulate
rapidity divergences through tilting light-cone directions also in the
case of real corrections.
\\

From a technical point, this does not imply any major
complications. However the real part of the 1-loop Green's function
Eq.~\eqref{eq:1} would receive a finite correction which in covariant
Feynman gauge is related to the square of the induced diagrams (last 2
diagrams in Fig.~\ref{fig:central}). As can be seen already at the
level of diagrams, such contributions arise as well for the
corresponding impact factors, which contain an identical diagram once
calculating the correction due to the emission of an additional real
gluon. As a consequence, it is straightforward to show that the
corresponding contributions cancel, once subtracted impact factor and
central contribution are being combined. Moreover, such a contribution
may be easily absorbed into a generalized version of the function
$\tilde{f}^{(1)}$, Eq.~\eqref{eq:2}. While including such
contributions does not provide any substantial complication, one deals
in that case with an entirely spurious contribution, which merely
arises due to our choice of our regulator and which has no physical
meaning. It seems therefore natural to employ a regulator which avoids
such a contribution altogether, at least at 1-loop. The modified
regulator is essentially identical to the previously used tilted
light-cone vectors, while the tilted elements are taken now to be
complex, {\it i.e.} we will use in the following
 \begin{align}
   \label{eq:ntilt}
   n^\pm \to n^{b,a} & = n^\pm + i e^{-\rho} n^\mp, & \rho \in \mathbb{R}.
 \end{align}
As a consequence one has for virtual corrections
\begin{align}
  \label{eq:n2}
  {n^{a,b}}^2 & = - 4 e^{- \rho},  & n^a \cdot n^b & = 2(1-e^{-2 \rho}) %& \frac{n^a  \cdot n^b}{{n^a}^2 {n^b}^2} = \sinh^2 \rho,
\end{align}
while real corrections yields
\begin{align}
  \label{eq:n2_real}
   |{n^{a,b}}|^2 & = 0,   & n^a \cdot (n^b)^* + c.c. & = 4(1+e^{-2 \rho}).
\end{align}
The spurious self-energy like contributions are therefore absent. The only disadvantage of this method is that terms of the form $\ln ({n^{a,b}}^2)$ in  virtual corrections can give  rise to undesired imaginary parts due to space-like $ {n^{a,b}}^2$. While at cross-section level such imaginary parts cancel naturally, if one limits oneself to NLO corrections, a consistent treatment of such contribution at amplitude level would require to absorb this imaginary part into the parameter $\rho$, {\it e.g.} through a suitable replacement $\rho \to \tilde{\rho} = \rho - i \pi /2$ etc. in the transition functions. \\

In the following calculation we will meet rapidity divergences which originate   both from high energy factorization and the QCD operator definition of the TMD gluon distribution and the corresponding soft factor, which we will consistently regulate through the tilting as described in Eq.~\eqref{eq:ntilt}, while we reserve the use of the regulator $\rho \to \infty$ for rapidity divergences due to high energy factorization. To be specific we define in the following the tilted Wilson lines of the TMD definition Eq.~\eqref{eq:TMDdef} and Eq.~\eqref{eq:Sy1y2_def} as
\begin{align}
  \label{eq:n1n2}
 n_{1,2}(y_{1,2}) & = n^\mp + i e^{- y_{1,2}} n^\pm,  \notag \\
  n(\sigma) & = n^- + i e^{-\sigma} n^+.
\end{align}
{}

\section{Determination of the gluon TMD}
\label{sec:gluonTMD}

\begin{figure*}[t]
 \includegraphics[width=0.8\textwidth]{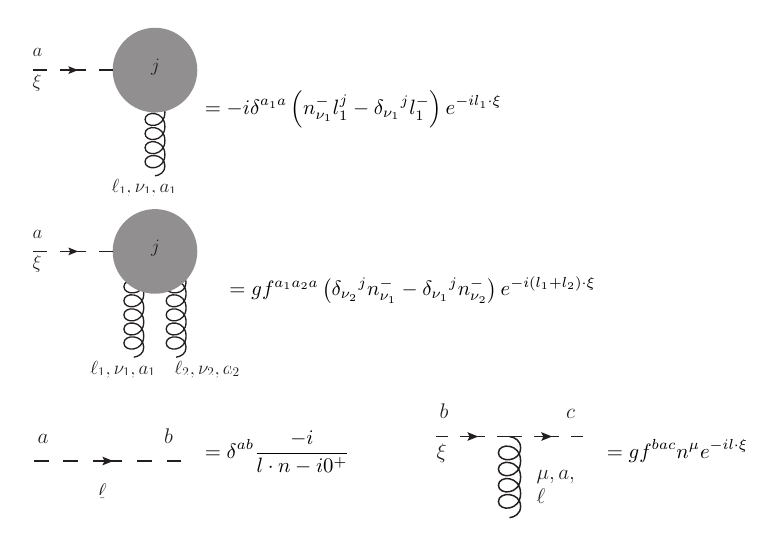}

  \caption{Feynman rules for the perturbative determination of the gluon TMD at amplitude level, {\it i.e.} for the evaluation of  $ \mathcal{W}^{n(\sigma)}_0 G^{-j}(0)] \big|r^b_+(k)\big \rangle$}
  \label{fig:feynTMD}
\end{figure*}

The goal of the following section is to determine the  gluon TMD Eq.~\eqref{eq:TMDdef} within high energy factorization, {\it i.e.} we aim at the determination of the following coefficient $C_{gg^*}$, implicitly defined through
\begin{align}
  \label{eq:first_def_Coeff}
    f_{g}&(\eta_a, \eta_b, y_c, \zeta_B, \qt, \mu) \notag \\ &  = \int \frac{d^2 \kt}{\pi} C^f_{gg^*} (\zeta_B, y_c, \eta_a, \qt, \kt, \mu) \mathcal{G}(\Delta \eta_{ab}, \eta_b; \kt), \notag \\
 h_{g}&(\eta_a, \eta_b, y_c, \zeta_B, \qt, \mu) \notag \\ & = \int \frac{d^2 \kt}{\pi} C^h_{gg^*} (\zeta_B, y_c, \eta_a, \qt, \kt, \mu) \mathcal{G}(\Delta \eta_{ab}, \eta_b; \kt).
\end{align}
Note that the TMD PDFs at first do not depend on the proton momentum fraction $x$, since high energy factorization requires to integrate over this longitudinal momentum fraction. Such a dependence therefore only arises through a special choice for the parameters $\eta_{a,b}$. To allow for a separate discussion of the different contributions of the gluon TMD,  we further define
\begin{widetext}
\begin{align}
  \label{eq:J}
  J^{ij}(q^-, \sigma, \qt, \mu) & =  \int \frac{d \xi^+ d^{2+2 \epsilon } {\bm \xi}}{2(2 \pi)^{3 + 2  \epsilon} p_B^-} 
e^{i q \cdot \xi } 
{\left \langle h(p_B)\left|\tr\left[\left(\mathcal{W}^{n(\sigma)}_\xi G^{-i}(\xi) \right)^\dagger \, \mathcal{W}^{n(\sigma)}_0 G^{-j}(0)\right] \right|h(p_B)\right \rangle} \bigg|_{\xi^- = 0}
\notag \\
\mathcal{S}(y_c, \sigma, \qt, \mu)  &= \int \frac{ d^{2 + 2 \epsilon} {\bm \xi}}{(2 \pi)^2} e^{i \qt_2 \cdot \xit} \, \,\tilde{ \mathcal{S}}(y_c, \sigma,\xit, \mu),
\end{align}  
\end{widetext}
where for the moment we define the TMD PDF in $2 + 2 \epsilon$ dimensions, since individual expressions are divergent and $q^- = x p_B^-$. We therefore obtain
\begin{align}
  \label{eq:Gamma_mmom}
  x\Gamma^{ij}(q^-, y_c, \qt, \mu)  = &\int d^{2 + 2 \epsilon} \qt_1   
                                     J^{ij}(q^-, \sigma, \qt_1, \mu)  \notag \\
  & \cdot \mathcal{S}((y_c, \sigma, \qt-\qt_1, \mu).
\end{align}
In the following we evaluate the above gluon TMD for an initial reggeized gluon state  with polarization $n^+$ at  1-loop. To be precise we consider
\begin{widetext}
\begin{align}
  \label{eq:Jhat}
    \bar{J}^{ij}(q^-, \sigma, \qt, \kt, \mu) & =  \int \frac{d \xi^+ d^{2+ 2 \epsilon } {\bm \xi}}{2(2 \pi)^{3 + 2  \epsilon} k^-} 
e^{i q \cdot \xi} \cdot \frac{1}{N_c^2-1}\sum_{b,b'} 
{\left \langle r^{b'}_+(k)\left|\tr\left[\left(\mathcal{W}^{n(\sigma)}_\xi G^{-i}(\xi) \right)^\dagger \, \mathcal{W}^{n(\sigma)}_0 G^{-j}(0)\right] \right|r^b_+(k)\right \rangle} \bigg|_{\xi^- = 0},
\end{align}  
\end{widetext}
where the reggeized gluon state $r^b(k)$ is defined with the normalization Eq.~\eqref{eq:dsig_hat}, appropriate for matching to collinear factorization in the limit of vanishing transverse momentum,{\it i.e.} 
\begin{align}
  \label{eq:porjection}
  v_\mu^a(\xi) \left | r^b_+(k) \right \rangle & = e^{i k \cdot \xi} n_\mu^+ \delta^{ab} \frac{k^-}{2 |\kt|} \bigg|_{k^+ = 0},
\end{align}
while high energy factorization requires  to integrate over the minus momentum, 
\begin{align}
  \label{eq:Jhatint}
   \hat{J}^{ij}(x, \qt_1, \kt) & = \int \frac{dk^-}{k^-}   \bar{J}^{ij}(x, \qt_1, \kt).
\end{align}
Feynman rules for the determination of perturbative corrections are summarized in Fig.~\ref{fig:feynTMD}. With the following convention to denote the perturbative expansion in $\alpha_s$ for a generic quantity $A(\alpha_s)$
\begin{align}
  \label{eq:expansion}
  A(\alpha_s) & = A^{(0)} + A^{(1)} ( \alpha_s) + \ldots,
\end{align}
where $A^{(n)} \sim \alpha_s^n$, we have finally at leading order
\begin{align}
  \label{eq:J0}
  \bar{J}^{ij, (0)} & = \frac{\qt^i \qt^j}{\qt^2}\delta^{(2 + 2 \epsilon)} (\qt - \kt) \delta \left(1 - \frac{k^-}{q^-} \right),
\notag \\
 \hat{J}^{ij, (0)} & = \frac{\qt^i \qt^j}{\qt^2}\delta^{(2 + 2 \epsilon)} (\qt - \kt),
\end{align}
and therefore
\begin{align}
  \label{eq:Cgg0}
  C_{gg^*}^{f, (0)}(\qt, \kt) & =   C_{gg^*}^{h, (0)}(\qt, \kt) =  \delta^{(2)}(\qt - \kt), 
\end{align}
and the TMD gluon distributions are up to an overall factor\footnote{The overall factor arises since -- at least at the level of bare distributions -- the unintegrated gluon density reduces to the collinear gluon in the double logarithmic limit after integration over $\int d \kt^2$ while the gluon TMD requires an integral over $\int d^2 \kt$, which gives rise to a relative factor of $\pi$.} of $1/\pi$ at leading order identical to the unintegrated gluon density  \cite{Dominguez:2011wm}
\begin{align}
  \label{eq:fho}
  f^{(0)} & = h ^{(0)}  =  \frac{1}{\pi} G\left(\Delta \eta_{ab}, \eta_b, \qt, \bar{f}^{(1)}\right),
\notag \\
\Gamma^{(0)ij} & = \frac{\qt^i \qt^j}{\qt^2 \pi} G\left(\Delta \eta_{ab}, \eta_b, \qt, \bar{f}^{(1)}\right),
\end{align}
where the $k_T$-factorization scheme defined in
Sec.~\ref{sec:kT_scale} yields expression which are closest to
conventional collinear factorization results.  Note that
the unintegrated gluon density is therefore directly related to the
operator definition of the gluon TMD. Moreover, in the dilute limit
{\it i.e.} considering only 2 reggeized gluon exchange, the
unintegrated gluon density is universal\footnote{Note that this
  universality breaks down, once corrections due to multiple reggeized
  gluon exchange are included see {\it e.g.}
  \cite{Altinoluk:2021ygv}}. We further stress that the distribution
of linearly polarized gluons in an unpolarized hadron is non-zero
within high energy factorization already at tree level, in contrast to
the result found within collinear factorization
\cite{Echevarria:2015uaa}. From a technical point of view this is of
course easily understood, since the initial gluon carries within high
energy factorization already finite $\kt$ and therefore gives rise to
such a distribution.

\subsection{One-loop calculation without soft-factor}
\label{sec:TMD1loop}

\begin{figure*}[t]
  \flushleft
  \parbox{1.9cm}{\includegraphics[height=3.2cm]{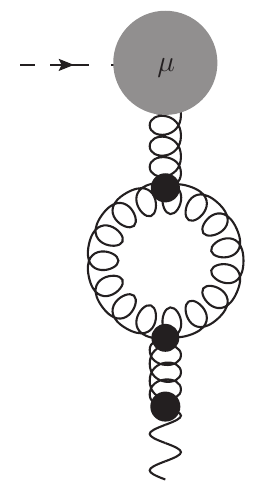}}
  \parbox{1.9cm}{\includegraphics[height=3.2cm]{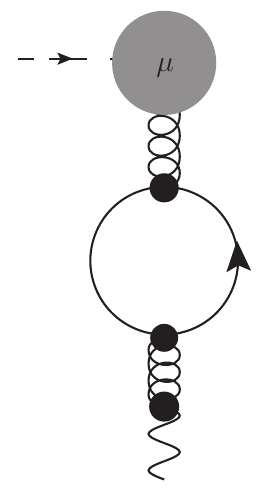}}
  \parbox{1.9cm}{\includegraphics[height=3.2cm]{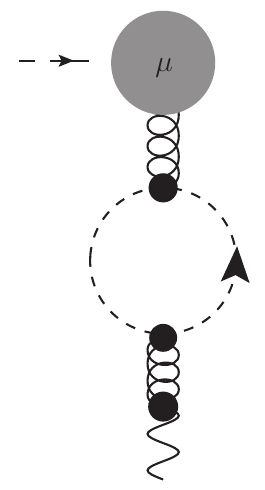}} 
  \parbox{1.9cm}{\includegraphics[height=3.2cm]{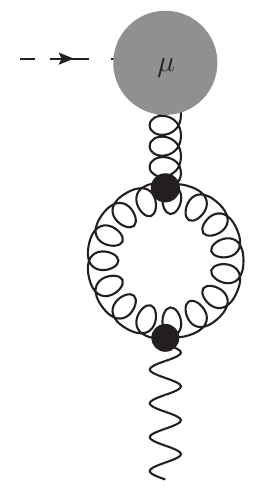}}
  \parbox{1.9cm}{\includegraphics[height=3.2cm]{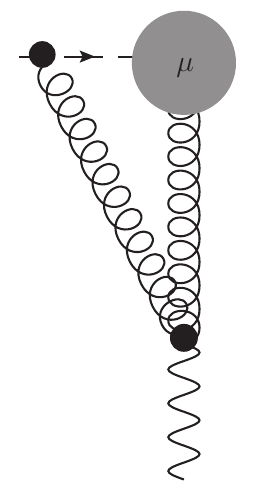}}
  \parbox{1.9cm}{\includegraphics[height=3.2cm]{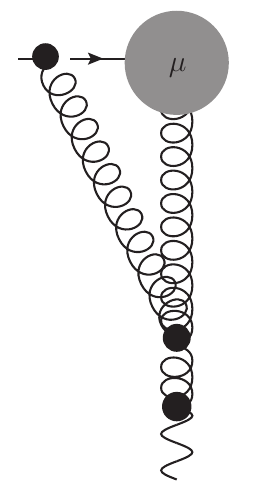}} 
 % \parbox{2cm}{\includegraphics[height=3cm]{v7.pdf}}
  \parbox{1.9cm}{\includegraphics[height=3.2cm]{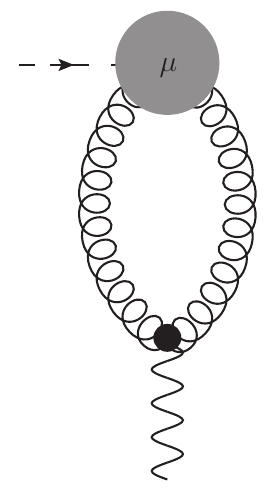}}
  \parbox{1.9cm}{\includegraphics[height=3.2cm]{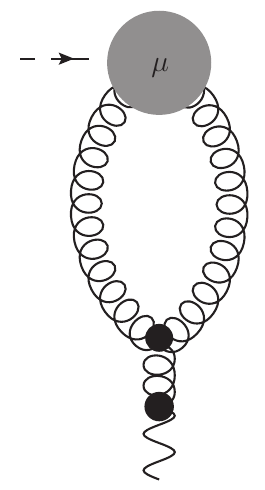}}
 \parbox{1.9cm}{\includegraphics[height=3.2cm]{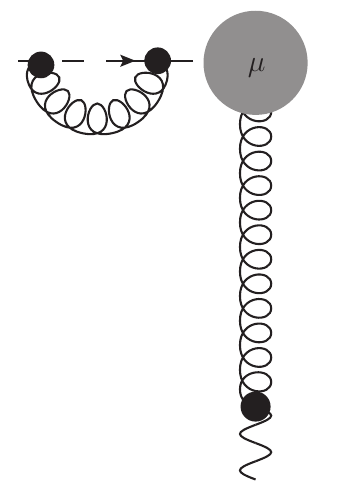}}
  %\parbox{1.5cm}{$\,$}
 % \parbox{2cm}{\includegraphics[height=3cm]{v11.pdf}}
 % \parbox{2cm}{\includegraphics[height=3cm]{v12.pdf}}
  \caption{Virtual corrections}
  \label{fig:virtual2}
\end{figure*}

\begin{figure*}[t]
  \centering
  \parbox{3cm}{\includegraphics[height=2.6cm]{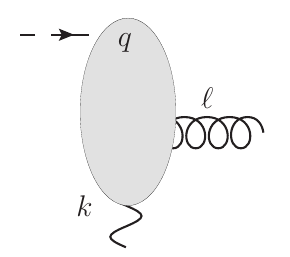}}
  \parbox{2.5cm}{\includegraphics[height=2.6cm]{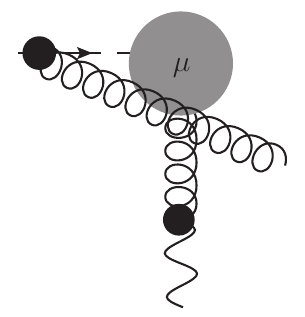}}
 \parbox{2.2cm}{\includegraphics[height=2.6cm]{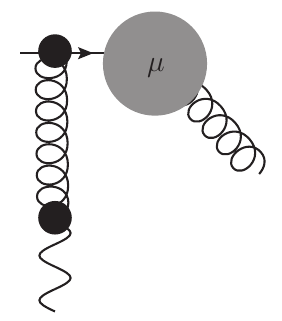}}
 \parbox{2.2cm}{\includegraphics[height=2.6cm]{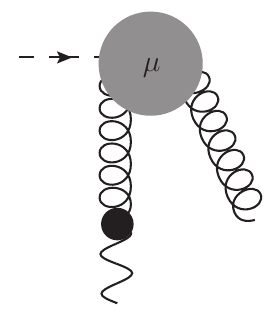}}
 \parbox{2.2cm}{\includegraphics[height=2.6cm]{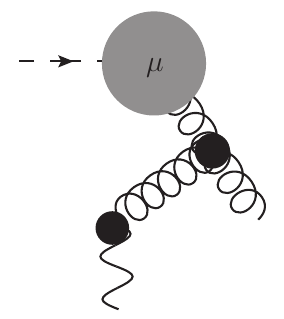}}
 \parbox{2.2cm}{\includegraphics[height=2.6cm]{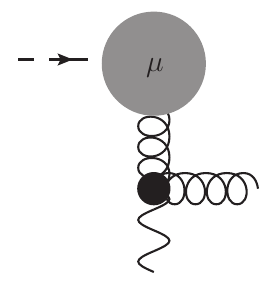}}
  \caption{Real corrections. Left: The reggeized gluon ($k$) is taken as incoming, the real gluon ($\ell$) and the momentum of the TMD PDF ($q$) as outgoing, $k = q + \ell$. Right: Contributing Feynman diagrams }
  \label{fig:real}
\end{figure*}

To regularize infrared and ultraviolet divergences we use dimensional
regularization in $d = 4 + 2 \epsilon$.  The virtual 1-loop correction
is provided by the set of diagrams Fig.~\ref{fig:virtual2}. The last
diagram in the second line vanish within dimensional regularization,
since it is scale-less. Including the contribution from the complex
conjugate amplitude, we obtain the following result
\begin{align}
  \label{eq:Jvirt}
 &  \bar{J}^{ij, (1)}_{\text{virt}}(q^-, \qt; k^-, \kt)  =   \delta \left(1 - \frac{k^-}{q^-} \right)  \hat{J}^{ij, (1)}_{\text{virt}}(q^-, \qt; \kt), \notag \\
&  \hat{J}^{ij, (1)}_{\text{virt}} (q^-, \qt;  \kt) =  \frac{\qt^i \qt^j}{\qt^2}\delta^{(2 + 2 \epsilon)} (\qt - \kt)\,  \left(\frac{\qt^2}{\mu^2} \right)^\epsilon   \notag \\
&  \cdot \,  \frac{\alpha_s C_A}{2 \pi} \bigg[\frac{1}{\epsilon^2} + %\notag %\\
%& \qquad +
   \frac{1}{\epsilon} \left(- \rho + \ln \frac{{q^-}^2}{\qt^2} - \frac{8}{3} + \frac{2 n_f}{3 C_A} \right) \notag \\
  & \hspace{2cm} + \frac{49}{9} - \frac{10 n_f}{9 C_A} - \frac{\pi^2}{3} \bigg] + \mathcal{O}(\epsilon)
, 
\end{align}
To obtain the projections on the distribution for the unpolarized TMD
PDF $f^g$ and the linearly polarized gluons $h^g$ in an unpolarized
hadron, we use
\begin{align}
  \label{eq:project}
  f_{g/B}(x, \qt) & = -g_{\perp, ij}  \Gamma^{ij}_{g/B}(x, \qt),
\notag \\
 h_{g/B}(x, \qt) & = \frac{2+ 2 \epsilon}{1 + 2 \epsilon}   \left(\frac{g^{ij}}{2 + 2 \epsilon} + \frac{\qt^i \qt^j}{\qt^2} \right)  \Gamma^{ij}_{g/B}(x, \qt),
\end{align}
 which amounts to replace the overall tensor
structure $\qt^i \qt^j/\qt^2$ by one for both the unpolarized and the
linearly polarized TMD. In particular, due to the presence of non-zero
initial transverse momentum, the virtual correction for both TMD
distributions are non-zero and agree with each other. Diagrams for real corrections are depicted in Fig.~\ref{fig:real}. Parameterizing $k^- = q^-/z$ and correspondingly $l^- =  q^- (1-z)/z$, a straightforward calculation yields
\begin{align}
  \label{eq:realcorrections}
  \bar{J}^{ij, (1)}_{\text{real}}(q^-, \qt; k^-, \kt) & =\frac{1}{\pi^{1 + \epsilon}} \int_0^1 dz   \delta \left(1 - \frac{z k^- }{q^-} \right)   \notag \\
  & \quad \frac{1}{(\qt - \kt)^2} \, \tilde{P}_{gg, r}^{(0),ij}(z, \qt, \kt), 
\end{align}  
where
\begin{widetext}
\begin{align}
  \label{eq:open}
  \frac{1}{(\qt - \kt)^2} P_{gg, r}^{(0)ij}&(z, \qt, \kt)  = \frac{\alpha_s C_A}{2 \pi \mu^{2 \epsilon} \Gamma(1-\epsilon)}
                                             \bigg[ \frac{-g_\perp^{ij}}{2} \frac{z(1-z) ((\qt - \kt)^2 - \qt^2)^2}{(z (\qt - \kt)^2 + (1-z) \qt^2) \kt^2} 
                                             +
                                             \frac{\qt^i \qt^j}{\qt^2} \bigg(\frac{2}{z [(\qt - \kt)^2  + e^{- \rho} \frac{(1-z)^2}{z^2} {q^-}^2 ]} \notag \\
  & - \frac{2}{z (\qt - \kt)^2 + (1-z) \qt^2} \bigg)
  +\frac{\kt^i \kt^{j}}{\kt^2  (\qt - \kt)^2} \left( \frac{2}{[(1-z) + e^{- \sigma} \frac{(\qt - \kt)^2 z^2}{(1-z)^2 {q^-}^2}]} - \frac{2  \qt^2}{\left[z (\qt - \kt)^2 + (1-z) \qt^2 \right]} \right) 
\notag \\
&
\hspace{1cm}
+ \frac{\kt^i \qt^{j} + \qt^i \kt^{j}}{ (\qt - \kt)^2} \frac{1}{\left[z (\qt - \kt)^2 + (1-z) \qt^2 \right] }
\bigg] + \mathcal{O}(e^{- \rho}, e^{- \sigma}),
\end{align}  
\end{widetext}
where we kept only track of those contributions of order $e^{- \rho}$ and $e^{-\sigma}$ which are needed to regulate integrals over the momentum fraction $z$ and/or transverse momenta, which are not convergent within dimensional regularization. Real splitting functions for the  unpolarized ($\tilde{P}_{gg}^{(f)}$) and linearly polarized  ($\tilde{P}_{gg}^{(h)}$) gluon read
\begin{widetext}
\begin{align}
  \label{eq:splitting_fordistributions_f}
    \frac{1}{\lt^2} \tilde{P}_{gg, r}^{(0)f}(z, \qt, \kt) & = \frac{\alpha_s C_A}{2 \pi \mu^{2 \epsilon} \Gamma(1-\epsilon)} \Bigg\{  \frac{z(1-z)(1 + \epsilon) (\lt^2 - \qt^2)^2}{\left[z \lt^2 + (1-z) \qt^2\right]^2 \kt^2}  
+ \frac{2}{\lt^2}\bigg[ 
\frac{1}{z \cdot \left[1  + e^{- \rho} \frac{(1-z)^2 {q^-}^2 }{z^2 \lt^2 } \right] }
\notag \\
         &+ 
 \frac{1}{(1-z)\cdot \left[1 + e^{- \sigma} \frac{\lt^2 z^2}{(1-z)^2 {q^-}^2}\right]} 
\bigg]
+ \frac{1}{ \lt^2} \frac{\kt^2 - 3(\qt - \kt)^2 - \qt^2}{\left[z (\qt - \kt)^2 + (1-z) \qt^2 \right] } 
\Bigg\}, \\
%\end{align}
%\begin{align}
 \label{eq:splitting_fordistributions_h}
  \frac{1}{\lt^2} \tilde{P}_{gg, r}^{(0)h}(z, \qt, \kt)  &= \frac{\alpha_s C_A}{2 \pi \mu^{2 \epsilon} \Gamma(1-\epsilon)} \Bigg\{
\frac{2}{z \lt^2 \cdot
 \left[1  + e^{- \rho} \frac{(1-z)^2 {q^-}^2 }{z^2 \lt^2 } \right] } 
 +     \frac{2}{(1-z)\lt^2\cdot \left[1 + e^{- \sigma} \frac{\lt^2 z^2}{(1-z)^2 {q^-}^2}\right]}\notag \\
  & \hspace{-1.5cm}\times \left(1+ \frac{4(1 +  \epsilon) \left( (\lt \cdot \qt)^2 - \lt^2 \qt^2 \right)}{(1 + 2 \epsilon)\lt^2 \qt^2 \kt^2} \right) 
   +
\frac{1 }{\left[z \lt^2 + (1-z) \qt^2 \right]} \left(\frac{(1 + 2\epsilon)2 \kt \cdot \qt + 2 \qt^2}{\lt^2 (1 + 2 \epsilon)} - \frac{2(1+ \epsilon)(\kt \cdot \qt)^2}{\kt^2 \lt^2 (1 + 2 \epsilon)} -2 \right)\Bigg\},
\end{align}  
\end{widetext}
where we used $\lt = \kt - \qt$.  Note that the splitting function
corresponding to $f_g$ coincides with the real transverse momentum
splitting function derived in \cite{Hentschinski:2017ayz} if we set
$\rho = \infty = \sigma$.
As demonstrated in
\cite{Hentschinski:2017ayz}, this splitting function coincides with
the DGLAP splitting function in the limit $\kt \to 0$, reduces to the
real part of the leading order BFKL kernel in the limit $z \to 0$ and
yields the real leading order kernel of the CCFM equation in the limit
$\qt \to \kt$. The present calculation provides on the other hand an
opportunity to determine the still missing virtual contribution to
this splitting function. We further note  -- in agreement with the
result presented in \cite{Echevarria:2015uaa} --  that the coefficient of the
singularity $z \to 1$ vanishes for the polarized splitting, Eq.~\eqref{eq:splitting_fordistributions_h} in the
collinear limit $\kt \to 0$, after averaging additionally over the
azimuthal angle of the incoming gluon. For finite $\kt$ this singularity is on the other hand present and
requires a treatment similar to the case of the unpolarized gluon TMD
PDF. With the integrated real corrections -- relevant for the current
discussion which is based on high energy factorization -- we finally
have:
\begin{align}
  \label{eq:higenergysplitting}
\hat{J}^{ij, (1)}_{\text{real}}(q^-, \qt;  \kt) & = \int_0^1 dz  \,  \frac{1}{(\qt - \kt)^2} \, \frac{\tilde{P}_{gg, r}^{(0),ij}(z, \qt, \kt)}{\pi^{1 + \epsilon}},
\end{align}
with
\begin{widetext}
\begin{align}
  \label{eq:J_fordistributions_f}
\hat{J}^{(1)f}_{\text{real}}(q^-, \qt; \kt) & = \frac{\alpha_s C_A}{2 \pi \pi_\epsilon} \Bigg\{  \frac{\sigma + \rho }{\lt^2}
+
\int\limits_0^1 dz \Bigg[
\frac{z(1-z)(1 + \epsilon) (\lt^2 - \qt^2)^2}{\left[z \lt^2 + (1-z) \qt^2\right]^2 \kt^2} 
+ \frac{1}{ \lt^2} \frac{\kt^2 - 3(\qt - \kt)^2 - \qt^2}{\left[z (\qt - \kt)^2 + (1-z) \qt^2 \right] } \Bigg]
\Bigg\}, \\
 \label{eq:J_fordistributions_h}
 \hat{J}^{(1)h}_{\text{real}}(q^-, \qt; \kt) & = \frac{\alpha_s C_A}{2 \pi \pi_\epsilon} \Bigg\{ \frac{\sigma + \rho}{\lt^2}
 + \left(\sigma + \ln \frac{{q^-}^2}{\lt^2}\right) \frac{2(1 +  \epsilon) \left( (\lt \cdot \qt)^2 - \lt^2 \qt^2 \right)}{(1 + 2 \epsilon)\lt^2 \qt^2 \kt^2} 
\notag \\
& \hspace{.5cm}+
\int\limits_0^1 dz \frac{1 }{\left[z \lt^2 + (1-z) \qt^2 \right]} \left(\frac{(1 + 2\epsilon)2 \kt \cdot \qt + 2 \qt^2}{\lt^2 (1 + 2 \epsilon)} - \frac{4(1+ \epsilon)(\kt \cdot \qt)^2}{\kt^2 \lt^2 (1 + 2 \epsilon)} -2 \right)\Bigg\},
\end{align}  
\end{widetext}
where the remaining integrals over $z$ are finite and $\pi_\epsilon$ is defined in Eq.~\eqref{eq:somedef}. If inserted into Eq.~\eqref{eq:first_def_Coeff}, the convolution integral over the reggeized gluon momentum $\kt$ gives still rise to an infra-red singularity. A possibility to extract these singularities is the use of  a phase space slicing parameter, see {\it e.g.} \cite{Hentschinski:2014lma,Hentschinski:2014bra,Hentschinski:2014esa} for an example within the high energy effective action. While this is sufficient to demonstrate finiteness at a
formal level, the use of such phase space slicing parameters is in
general complicated for numerical studies at NLO accuracy. For the
case of collinear NLO calculation, the by now conventional tool to
overcome this difficulty are subtraction methods, in
particular  dipole subtraction 
\cite{Catani:1996vz}. In \cite{Hentschinski:2020tbi}  this has been slightly generalized and applied to the case of divergences which arise due to  convolution integrals of transverse momenta. In particular the  following decomposition has been proposed:
\begin{align}
  \label{eq:subtR2}
   \int &\frac{d^{2+2\epsilon} \lt}{\pi^{1+\epsilon}} 
   \,\frac{\kappa(\lt)}{\lt^2}\, 
   G\left((\qt + \lt)^2\right)  \notag \\
& = 
  \int\frac{d^{2} \lt}{\pi }
   \left[\frac{\kappa(\lt)}{\lt^2}\right]_+G((\qt + \lt)^2) 
\notag\\& 
  +\int \frac{d^{2+2\varepsilon}\lt}{\pi^{1+\varepsilon}}
   \,\frac{\kappa(\lt)}{\lt^2}\,
   \frac{\qt^2G(\qt^2)}{\lt^2 + (\qt + \lt)^2} \;+\; \mathcal{O}(\epsilon)
\;\;, 
\end{align}
with
\begin{align}
  \label{eq:plusrp}
   \int \frac{d^{2} \lt}{\pi} &
   \left[\frac{\kappa(\lt)}{\lt^2} \right]_+
   G((\qt + \lt)^2) 
 \equiv 
\int\frac{d^{2}\lt}{\pi}
                       \,\frac{\kappa(\lt)}{\lt^2}\, \notag \\
  & \left[G((\qt + \lt)^2) - \frac{\qt^2 G(\qt^2)}{ \,\lt^2 + (\qt + \lt)^2} \right]
\;\;.
\end{align}
The expression in the squared brackets on the right-hand side vanish in the limit $|\lt| \to 0$, and $G(\kt)$ is a function which parameterizes the transverse momentum dependence of the reggeized gluon state.
The function $\kappa(\lt)$ is such that the integral on the right hand side of \eqref{eq:plusrp} is well-defined, which in practice means that it does not behave worse than $\ln|\lt|$ for $|\lt|\to0$ and $|\lt|\to\infty$.
 Furthermore, it should be such that the integral in the second line of \eqref{eq:subtR2} can be calculated analytically. Note that the factor $\qt^2/ [\lt^2 + (\qt + \lt)^2]$ is needed to achieve convergence in the ultraviolet. In the current setup we merely require the case  $\kappa(\lt)=1$ with
\begin{align}
  \label{eq:int1}
  \int \frac{d^{2 + 2 \epsilon}\lt}{\pi^{1 + \epsilon}}
  \frac{1}{\lt^2 [\lt^2 + (\qt + \lt)^2]} & %= \frac{\Gamma(1-\epsilon)\Gamma^2(\epsilon)}{2 \Gamma (2\epsilon)} (\qt^2)^{\epsilon -1}
  =
  \frac{\Gamma(1-\epsilon)}{\epsilon  (\qt^2)^{1-\epsilon}}  + \mathcal{O}(\epsilon),
%\notag \\
% \int \frac{d^{2 + 2 \epsilon}\lt}{\pi^{1 + \epsilon}}
%  \frac{-\ln\lt^2}{\lt^2 [\lt^2 + (\qt + \lt)^2]}
% & = 
%  \frac{\Gamma(1-\epsilon)}{(\qt^2)^{1 - \epsilon}} \left(\frac{1}{\epsilon^2} - \frac{\ln \qt^2}{\epsilon} - \frac{\pi^2}{4} \right) + \mathcal{O}(\epsilon)
%\notag\\
%  \int \frac{d^{2 + 2 \epsilon} \lt}{\pi^{1 + \epsilon}} 
%  \frac{\left(\qt\cdot \lt\right)^2}{(\lt^2)^2[\lt^2+(\qt + \lt)^2]} 
%  & = 
%  \frac{\Gamma(1-\epsilon)}{(\qt^2)^{-\epsilon}} \left(\frac{1}{2 \epsilon} - 1 + \ln 2 \right) + \mathcal{O}(\epsilon) .
\end{align}
and
\begin{widetext}
\begin{align}
  \label{eq:J_fordistributions_reg}
\hat{J}^{(1)f}_{\text{real}}(q^-, \qt; \kt) & = \frac{\alpha_s C_A}{2 \pi} \Bigg\{ (\rho + \sigma)\left(\delta^{(2)}(\lt) \frac{1}{\epsilon} \left(\frac{\qt^2}{\mu^2}\right)^\epsilon + \frac{1}{\pi^{1 + \epsilon}}\left[\frac{1}{\lt^2} \right]_+ \right) 
\notag \\
& +
\frac{1}{\pi^{1 + \epsilon}} \int\limits_0^1 dz \Bigg[
\frac{z(1-z) (\lt^2 - \qt^2)^2}{\left[z \lt^2 + (1-z) \qt^2\right]^2 \kt^2} 
%\notag \\
%&
%\hspace{6cm}
+ \frac{1}{ \lt^2} \frac{\kt^2 - 3(\qt - \kt)^2 - \qt^2}{\left[z (\qt - \kt)^2 + (1-z) \qt^2 \right] } \Bigg]
\Bigg\} + \mathcal{O}(\epsilon), \\
 \label{eq:J_fordistributions_h_reg}
 \hat{J}^{(1)h}_{\text{real}}(q^-, \qt; \kt) & = \frac{\alpha_s C_A}{2 \pi} \Bigg\{ 
 (\rho + \sigma)\left(\delta^{(2)}(\lt) \frac{1}{\epsilon} \left(\frac{\qt^2}{\mu^2}\right)^\epsilon + \frac{1}{\pi^{1+ \epsilon}}\left[\frac{1}{\lt^2} \right]_+ \right) 
\notag \\
& 
 + \left(\sigma + \ln \frac{{q^-}^2}{\lt^2}\right) \frac{2 \left( (\lt \cdot \qt)^2 - \lt^2 \qt^2 \right)}{\pi^{1+ \epsilon}\lt^2 \qt^2 \kt^2} 
+
\frac{1}{\pi^{1 + \epsilon}} \int\limits_0^1 dz \frac{(\qt^2 - \lt^2)2 \kt \cdot \lt }{\left[z \lt^2 + (1-z) \qt^2 \right] \lt^2 \kt^2} \Bigg\} + \mathcal{O}(\epsilon),
\end{align}  
\end{widetext}
where the corresponding convolution integral can now be defined in $d = 2$ dimensions.

\subsection{Soft factor, counter terms  and renormalization}
\label{sec:softrenom}

The above result contains both rapidity divergences due to high energy factorization $(\rho \to \infty)$ and the TMD definition ($\sigma \to \infty$) as well as single and double poles in $1/ \epsilon$. Rapidity divergences due to high energy factorization require the  subtraction of high energy factorized contributions to the above correlator as well as the application of the transition function, as given by Eq.~\eqref{eq:4}.  Rapidity divergences due to soft radiation require the soft factor Eq.~\eqref{eq:softC11} which we did not include so far.   With the 1-loop expansion
\begin{align}
  \label{eq:Sy1y2}
  {S}&(y_1, y_2; \qt_2)  = \delta^{(2 + 2 \epsilon)}(\qt_2)
       \notag \\
  & + \frac{\alpha_s C_A}{2 \pi \pi_\epsilon} (y_1 + y_2)\frac{1 + e^{-y_1 - y_2}}{1 - e^{-y_1 - y_2}} \frac{1}{\qt_2^2} + \mathcal{O}(\alpha_s^2),
\end{align}
and  taking the limit\footnote{See \cite{Collins:2012uy} and \cite{Echevarria:2012js} for a detailed discussion of these limits} $\sigma, y_n  \to \infty$, we  obtain  finally for the soft function in  momentum space  at 1-loop
\begin{align}
  \label{eq:softfunc_mom}
  \mathcal{S}(2y_c, \sigma; \qt_2) & = \delta^{(2 + 2 \epsilon)}(\qt_2) - \frac{\alpha_s C_A (\sigma - 2 y_c)}{2 \pi \pi_\epsilon \qt_2^2}  + \mathcal{O}(\alpha_s^2),
\end{align}
where $y_c$ is a finite rapidity and takes a role related to a factorization scale, similar to the parameter $\eta_a$ in the case of high energy factorization; in particular it enters directly the defintion of the scales $\zeta_{A,B}$ defined in Eq.~\eqref{eq:zeta_scales}. The 1-loop contribution to the TMD gluon densities due to the soft function is finally given by
\begin{align}
\label{eq:softtoTMD}
    \hat{\Gamma}^{ij(1)}_{soft}(x, \qt, \kt) & = - \frac{\alpha_s C_A (\sigma - 2 y_c) }{2 \pi \pi_\epsilon (\qt - \kt)^2} \frac{\kt^i \kt^j}{\kt^2 }+ \mathcal{O}(\alpha_s^2).
\end{align}
 Note that  the 1-loop soft-function, which  consists of its real part only within dimensional regularization,  agrees with the real part of the 1-loop BFKL kernel. We  believe that this coincidence is limited to the 1-loop case and does not hint at a general universality of the rapidity dependence of the soft function. In particular, at the diagrammatic level the soft-function does not give rise to the complete Lipatov vertex, but only to terms related to the induced contributions,  see also the discussion in \cite{Rothstein:2016bsq} in the context of soft-collinear effective theory for high energy scattering. \\

 As a last step we need to combine virtual (Eq.~\eqref{eq:Jvirt}  and real (Eqs.~\eqref{eq:J_fordistributions_f}\eqref{eq:J_fordistributions_h}) corrections, with the soft factor, making use of the   appropriate projections. Note that the contribution of the soft-factor merely amounts to a replacement of the regulator  $\sigma$ by the factorization parameter $y_c$.  We obtain for the 1-loop high energy subtracted and renormalized 1-loop coefficients $\tilde{C}^{(1)f,h}$ the following result:
 \begin{widetext}
\begin{align}
  \label{eq:J_fordistributions_ren}
\tilde{C}^{(1)f}(q^-, \qt; \kt)  = \frac{\alpha_s C_A}{2 \pi} &\Bigg\{ \delta^{(2)}(\lt) \bigg[
                                  \frac{1}{\epsilon^2} + \frac{1}{\epsilon} \left( \ln \frac{(q^-)^2 e^{2 y_c}}{\mu^2} - \frac{\beta_0}{2 C_A} \right) + \frac{67}{18} - \frac{5}{9} - \frac{\pi^2}{3}
                                  +  \ln \frac{\qt^2}{\mu^2} \left( \ln \frac{(q^-)^2 e^{2 y_c}}{\mu^2} - \frac{\beta_0}{2 C_A} \right) \notag \\
  & - \frac{1}{2} \ln^2 \frac{\qt^2}{\mu^2}  \bigg]
 +  2(y_c- \eta_a)  \frac{1}{\pi^{1 + \epsilon}}\left[\frac{1}{\lt^2} \right]_+  
 +
\frac{1}{\pi^{1 + \epsilon}} \int\limits_0^1 dz \Bigg[
\frac{z(1-z) (\lt^2 - \qt^2)^2}{\left[z \lt^2 + (1-z) \qt^2\right]^2 \kt^2} 
\notag \\ &
    +
 \frac{1}{ \lt^2} \frac{\kt^2 - 3(\qt - \kt)^2 - \qt^2}{\left[z (\qt - \kt)^2 + (1-z) \qt^2 \right] } \Bigg]
\Bigg\}
  +
 \bar{f}^{(1)}(\qt, \kt)+ \mathcal{O}(\epsilon), \\
%\end{align}
% \begin{align}
\label{eq:J_fordistributions_h_ren}
\tilde{C}^{(1)h}(q^-, \qt; \kt)  = \frac{\alpha_s C_A}{2 \pi } &\Bigg\{ 
\delta^{(2)}(\lt)  \bigg[
\frac{1}{\epsilon^2} + \frac{1}{\epsilon} \left( \ln \frac{(q^-)^2 e^{2 y_c}}{\mu^2} - \frac{\beta_0}{2 C_A} \right) + \frac{67}{18} - \frac{5}{9} - \frac{\pi^2}{3}
                                                                 +  \ln \frac{\qt^2}{\mu^2} \left( \ln \frac{(q^-)^2 e^{2 y_c}}{\mu^2} - \frac{\beta_0}{2 C_A} \right)
  \notag \\ &- \frac{1}{2} \ln^2 \frac{\qt^2}{\mu^2}  \bigg] 
 +  
2(y_c- \eta_a)  \frac{1}{\pi^{1 + \epsilon}}\left[\frac{1}{\lt^2} \right]_+  
+ 
           2 \left(y_c + \ln \frac{{q^-}}{|\lt|}\right) \frac{2 \left( (\lt \cdot \qt)^2 - \lt^2 \qt^2 \right)}{\pi^{1+ \epsilon}\lt^2 \qt^2 \kt^2} 
\notag \\
& \hspace{2cm} 
+ 
\frac{1}{\pi^{1 + \epsilon}} \int\limits_0^1 dz \frac{(\qt^2 - \lt^2)2 \kt \cdot \lt }{\left[z \lt^2 + (1-z) \qt^2 \right] \lt^2 \kt^2} \Bigg\}
  + \bar{f}^{(1)}(\qt, \kt) + \mathcal{O}(\epsilon),
\end{align}
   \end{widetext}
where
\begin{align}
  \label{eq:beta0}
  \beta_0 & = \frac{11 C_A}{3} - \frac{2 n_f}{3}.
\end{align}
While the above expression no longer carries rapidity divergences, it  still comes with several poles  in $1/\epsilon$, which are of ultraviolet origin and which require renormalization. The corresponding renormalization constant is identical for both unpolarized and linearly polarized gluons and is  obtained as
\begin{align}
  \label{eq:renomfac}
  \mathcal{Z}_{G} & = 1  - \frac{\alpha_s C_A}{2 \pi} \left[\frac{1}{\epsilon^2} + \frac{1}{\epsilon} \left(  \ln \frac{\zeta}{\mu^2} -\frac{\beta_0}{2C_A}\right) \right],
\end{align}
which gives rise to the following anomalous dimension
\begin{align}
  \label{eq:anom}
  \gamma_{G} \left(\alpha_s(\mu), \ln \frac{\zeta}{\mu^2} \right)  &= \frac{d \ln  \mathcal{Z}_{G}}{d \ln \mu} \notag \\ & =  \frac{\alpha_s}{2 \pi} \left[\beta_0 - 2 C_A \ln \frac{\zeta}{\mu^2}   \right],
\end{align}
where we used $d \alpha_s/d \ln \mu = 2 \epsilon \alpha_s $ and $\zeta  = (q^-)^2 e^{2 y_c}$.  Note
that the above anomalous dimension agrees with the corresponding
result obtained within a treatment based on collinear factorization
\cite{Echevarria:2015uaa}. This is indeed to be expected since it
arises due to the renormalization of ultraviolet divergences, which
are naturally independent of the non-zero transverse momentum of the
initial state gluon. We however stress  that the linearly polarized
TMD gluon distribution does not give rise to the above anomalous
dimension within collinear factorization, since
the corresponding distribution vanishes within collinear factorization
at tree-level; the 1-loop result is therefore not
renormalized. We finally obtained for the renormalized coefficients
\begin{widetext}
\begin{align}
  \label{eq:J_fordistributions_reg2}
\hat{C}^{(1)f}_{gg*}&(\zeta_B, y_c, \eta_a, \qt, \kt, \mu, \bar{f}^{(1)})  = \frac{\alpha_s C_A}{2 \pi} \Bigg\{ \delta^{(2)}(\lt) \bigg[
 \ln \frac{\qt^2}{\mu^2} 
\left( \ln \frac{\zeta_B}{\mu^2} - \frac{\beta_0}{2 C_A} \right)
 - \frac{1}{2} \ln^2 \frac{\qt^2}{\mu^2}
%\notag \\
%&
  + \frac{67}{18} - \frac{5 n_f}{9 C_A} - \frac{\pi^2}{3} \bigg]
 \notag \\
&
+ 
 2(y_c- \eta_a)  \frac{1}{\pi^{}}\left[\frac{1}{\lt^2} \right]_+  
 +
\frac{1}{\pi^{}} \int\limits_0^1 dz \Bigg[
\frac{z(1-z) (\lt^2 - \qt^2)^2}{\left[z \lt^2 + (1-z) \qt^2\right]^2 \kt^2}  
%\notag \\
%&
%\hspace{2cm}
+ \frac{1}{ \lt^2} \frac{\kt^2 - 3(\qt - \kt)^2 - \qt^2}{\left[z (\qt - \kt)^2 + (1-z) \qt^2 \right] } \Bigg]
\Bigg\} +
 \bar{f}^{(1)}(\qt, \kt) 
\\
 \label{eq:J_fordistributions_h_reg2}
\hat{C}^{(1)h}_{gg*}&(\zeta_B, y_c, \eta_a, \qt, \kt, \mu,  \bar{f}^{(1)})  = \frac{\alpha_s C_A}{2 \pi } \Bigg\{ 
\delta^{(2)}(\lt) \bigg[ \ln \frac{\qt^2}{\mu^2} \left[ \ln \frac{\zeta_B}{\mu^2} - \frac{\beta_0}{2 C_A} \right] - \frac{1}{2} \ln^2 \frac{\qt^2}{\mu^2}
%\notag \\
%&
+  \frac{67}{18} - \frac{5n_f}{9 C_A} - \frac{\pi^2}{3}
 \bigg] 
\notag \\
& %\hspace{-1.4cm}
 +  
2(y_c- \eta_a)  \frac{1}{\pi^{}}\left[\frac{1}{\lt^2} \right]_+  
+ 
            \left(y_c + \ln \frac{{q^-}}{|\lt|}\right) \frac{4 \left( (\lt \cdot \qt)^2 - \lt^2 \qt^2 \right)}{\pi^{}\lt^2 \qt^2 \kt^2} %\notag \\
%& \hspace{2cm} 
+ 
\frac{1}{\pi^{}} \int\limits_0^1 dz \frac{(\qt^2 - \lt^2)2 \kt \cdot \lt }{\left[z \lt^2 + (1-z) \qt^2 \right] \lt^2 \kt^2} \Bigg\} + \bar{f}^{(1)}(\qt, \kt)
 %+ \mathcal{O}(\epsilon),
\end{align}  
\end{widetext}
where we made now the dependence on various  parameters $y_c$ and $\eta_a$  explicit. The above expressions for the 1-loop cofficients of unpolarized and linearly polarized gluon TMD are one of the main results of this work.

\section{Evolution}
\label{sec:evolution}

The coefficients Eqs.~\eqref{eq:J_fordistributions_reg2},~\eqref{eq:J_fordistributions_h_reg2} depend on three factorization scales
and/or parameters: $\mu$ (renormalization scale), $y_c$ (evolution
parameter of the soft function) and $\eta_a$ (evolution parameter of
the unintegrated gluon density). In addition we still have the dependence on
the function $\bar{f}^{(1)}$, which is also related to the unintegrated gluon density. The dependence on the renormalization
scale and the factorization parameter $y_c$ gives rise to
CSS resummation framework, 
\cite{Collins:1981uw,Collins:1981uk,Collins:1984kg}.  In the treatment
established for collinear initial states, see {\it e.g.}
\cite{Collins:2011zzd,Echevarria:2015uaa,Aybat:2011zv} it is customary
to consider to this end the Fourier-transform of the TMD coefficient to transverse
coordinate space
\begin{align}
  \label{eq:FT}
  &  \tilde{C}^{(1)i}_{gg*}\left(\zeta_B, y_c, \eta_a, \bt, \kt, \mu, \bar{f}^{(1)}\right) \notag \\
  & = \int d^2 \qt\, e^{ i \qt \cdot \bt}\, \hat{C}^{(1)i}_{gg*}\left(\zeta_B, y_c, \eta_a, \qt, \kt, \mu, \bar{f}^{(1)}\right), 
\end{align}
where $i=f,h$ and then to evolve the coefficient in coordinate space,
\begin{widetext}
\begin{align}
  \label{eq:evolution}
    \tilde{C}^{(1)i}_{gg*}\left(\zeta_{B,f}, \ln \frac{\sqrt{\zeta_{B,f}}}{q^-}, \eta_a, \bt, \kt, \mu_f, \bar{f}^{(1)}\right) 
 = \tilde{R}\left(\bt; \zeta_{B,f}, \mu_f,\zeta_{B,i}, \mu_i\right)  \tilde{C}^{(1)i}_{gg*}\left(\zeta_B, \ln \frac{\sqrt{\zeta_{B,i}}}{q^-}, \eta_a, \bt, \kt, \mu_i, \bar{f}^{(1)}\right),
\end{align}  
\end{widetext}
with the evolution operator
\begin{align}
  \label{eq:evolution_kernelCSS}
  \tilde{R}& \left(\bt; \zeta_{B,f}, \mu_f,\zeta_{B,i}, \mu_i\right)= \notag \\ & = \exp \left\{\int\limits_{\mu_i}^{\mu_f} \frac{d \bar{\mu}}{\bar{\mu}} \gamma_G\left(\alpha_s(\bar{\mu}), \ln \frac{\zeta_{B,f}}{\bar{\mu}^2} \right) \right\}
   \notag \\ & \hspace{3cm}  \times
                \left(\frac{\zeta_{B,f}}{\zeta_{B,i}} \right)^{-\frac{\tilde{K}_{CS}(\bt, \mu_i)}{2} }.
\end{align}
In the above expression, the CS kernel $\tilde{K}_{CS}(\bt, \mu_i)$ is in general assumed to have
both perturbative and non-perturbative contributions; for a detailed
discussion see
\cite{Collins:2011zzd,Echevarria:2015uaa,Aybat:2011zv}. The
non-perturbative contribution arises due to the 
inverse Fourier transform, which requires  to integrate over large
values of $\bt$, well into the non-perturbative region. Note that a similar
statement applies in principle for the convolution integral of
Eq.~\eqref{eq:kT_def_Coeff}, if the unintegrated gluon distribution
does not drop-off sufficiently fast for small values of transverse
momentum $\kt$. The actual evolution takes place in two steps: first
one evolves the coefficient at a certain initial renormalization scale
$\mu_i$ from an initial rapidity $y_{c,i}$ -- parameterized through
$\zeta_{B,k} = (q^-)^2 e^{2 y_{c,k}}$, $k=i,f$ -- to a final rapidity
$y_{c,f}$. The second step evolves then the TMD PDF from the initial
to the final renormalization scale. While the value of
the final renormalization scale is of the order of the hard scale,
{\it i.e.} the Higgs mass for the current example, the initial
renormalization scale $\mu_i$ must be chosen such that it minimizes
the perturbative correction to the TMD coefficients. In collinear calculations it it is naturally taken to be of the order of the transverse momentum $\qt$ or its inverse conjugate coordinate $\mu_b = 2 e^{-2 \gamma_E}/|\bt|$ with $\gamma_E \simeq 0.577216 $ the Euler constant.  In the current  setup,  the optimal choice is far from apparent, since the coefficients depend on multiple scales due to non-zero initial transverse momenta. In particular, the transverse momenta $\kt$ and $\qt$ are at least at first not necessarily of the same order of magnitude.

\subsection{A comparison of the kernels of CS and BFKL evolution}
\label{sec:comparison}

Both CS and BFKL evolution describe evolution in rapidity. It is therefore natural to expect that both evolution and their respective kernels have a certain overlap. For the derivation of the CS kernel, we follow closely \cite{Collins:2011zzd, Aybat:2011zv}, where the kernel of the Collins-Soper evolution equation is defined through the $y_c$ dependence of the renormalized soft factor,
\begin{align}
  \label{eq:defCS}
   \tilde{K}_{\text{CS}}(\xit, \mu) & = \frac{\partial}{\partial y_c}\ln\left[ \tilde{\mathcal{S}}(\xit, y_c, \mu) \mathcal{Z}_G(y_c)\right],
\end{align}
and is itself subject to the following renormalization group equation,
\begin{align}
  \label{eq:RG_for_CS}
  \frac{d \tilde{K}_{\text{CS}}}{d \ln \mu^2} & = - \Gamma_{\text{cusp}}^A (\alpha_s(\mu)), \quad 
\Gamma_{\text{cusp}}^A= \sum_{n=1}^\infty \left(\frac{\alpha_s}{4\pi}\right)^n \Gamma^A_{n-1}, \notag \\  & \hspace{-1.2cm}\Gamma_0^A  = 4C_A, \quad \Gamma_1^A =  \Gamma_0^A\left[\left(\frac{67}{9} - \frac{\pi^2}{3}\right) C_A-\frac{10 n_f}{9} \right],
\end{align}
where $\Gamma_{\text{cusp}}^A$ is the cusp anomalous dimension in the adjoint representation, see \cite{Idilbi:2005ni,Idilbi:2005er,Echevarria:2015uaa} for higher order terms. At 1-loop one finds 
\begin{align}
  \label{eq:CSkernel}
  \tilde{K}_{CS}^{(1)}(\xit, \mu) & = \frac{\alpha_s C_A}{\pi} \ln \left(\frac{4 e^{-2 \gamma_E}}{\xit^2 \mu^2} \right).
\end{align}
While the representation in transverse coordinate space is very useful for a direct solution of the CS-equation through exponentiation of the CS kernel as in Eq.~\eqref{eq:evolution_kernelCSS}, it is also instructive to formulate the CS-equation in momentum space, which allows for a  direct comparison with the BFKL equation. In particular, defining in complete analogy to the BFKL treatment a CS Green's function $ G_{i}(\Delta y,\kt_1, \kt_2)$, $i=$BFKL, CS,  such that 
\begin{align}
  \label{eq:Greens_functions}
  G_{i}(0,\kt_1, \kt_2)& = \delta^{(2)}(\kt_1 - \kt_2), \qquad i = \text{BFKL, CS} \notag \\
  \frac{d G_{i}(\Delta y ,\kt_1, \kt_2)}{d \Delta y}   & = \int \frac{d^2 \kt}{\pi} K_i(\kt_1, \kt)  G_{i}(\Delta y ,\kt, \kt_2),
\end{align}
one finds at 1-loop the following simple relation between both kernels:
\begin{align}
  \label{eq:relationKernels}
  K_{\text{CS}}^{(1)}(\kt_1, \kt_2, \mu) & =   K_{\text{BFKL}}^{(1)}(\kt_1, \kt_2) \notag \\
  & - \frac{\alpha_s C_A}{\pi} \delta^{(2)}(\kt_1 - \kt_2) \ln \frac{\mu^2}{\kt^2_1}.
\end{align}
  The presence of this factor  can be explained as follows. As it is well known, the virtual correction to the 1-loop BFKL kernel is directly related to the  gluon Regge trajectory $\omega(\epsilon, \kt)$ which in transverse momentum space can be written as
\begin{align}
  \label{eq:traj}
  \omega(\epsilon, \kt^2) & =- \frac{\alpha_s C_A}{4 \pi \mu^{2 \epsilon} \Gamma(1-\epsilon)} \int \frac{d^{2 + 2 \epsilon} \lt}{\pi^{1+ \epsilon}} \frac{\kt^2}{\lt^2 (\kt - \lt)^2} \notag \\
  & = -\frac{\alpha_s C_A}{2 \pi} \frac{1}{ \epsilon} \left(\frac{\kt^2}{\mu^2} \right)^\epsilon.
\end{align}
 The CS equation is on the other hand limited to soft radiation, which implies restriction to  transverse momenta  $|\lt| \ll |\kt|$ and  $|\lt - \kt| \ll |\kt|$. The  integrand in the above expression therefore reduces to
\begin{align}
  \label{eq:reducedintegral}
  \frac{\kt^2}{\lt^2 (\kt - \lt)^2}& \simeq \frac{1}{\lt^2} + \frac{1}{(\lt - \kt)^2},
\end{align}
and the integral vanishes  within dimensional regularization through a cancellation of the infra-red and ultra-violet $1/\epsilon$ poles. Removing on the other hand the UV pole through  renormalization, one finally ends up with the virtual contribution to the CS kernel. The CS evolution can be therefore understood as the soft approximation to the complete BFKL kernel. Indeed, such an identification is natural, since the CS kernel arises from the rapidity divergence of the soft-function, while the BFKL kernel from the rapidity divergence of partonic cross-sections in the high energy limit $\sqrt{s} \to \infty$. \\

The above discussion clearly suggests that the CS and BFKL evolution
are closely related to each other, with the  CS kernel as the soft limit of the complete BFKL
kernel. When considering  CS and BFKL evolution for the same quantity, it is
therefore  necessary to remove the overlap of both evolution
equations or alternatively to restrict them to distinct regions in phase space, which are then covered by the regarding evolution equation. In particular the various evolution parameters must  obey the following ordering
\begin{align}
  \label{eq:ordering}
  y_c^f > y_c^i &= \eta_a > \eta_b;
\end{align}
it is needed to separate the phase space covered by BFKL evolution
(rapidity evolution of the entire cross-section) and CS evolution
(rapidity evolution of soft gluons only). Clearly, soft gluons form a
sub-set of the complete cross-section and cannot be evolved separately
from the latter in rapidity.

\subsection{ $k_T$ factorization and alternative schemes}
\label{sec:CSSevolvKT}

In the following we investigate our result for a specific choice of the evolution parameter of the unintegrated gluon density ,which we fix to coincide with the proton momentum fraction $x$, {\it i.e.} we consider now the equivalent of Eq.~\eqref{eq:first_def_Coeff}, but with the following choices for the parameters of the high energy evolution, following the results of Sec.~\ref{sec:kT_scale}
\begin{align}
  \label{eq:fixKTparameters}
  \Delta \eta_{ab} & = \ln \frac{x_0}{x},  \quad \eta_a  = \ln \frac{x_0 M}{q^-}, \quad  \eta_b =  \ln \frac{ M}{p_b^-}, \notag \\
\bar{f}^{(1)} (\qt, \kt) & = \ln\left( \frac{x_0 M}{|\qt|}\right) \frac{1}{\pi} \left[\frac{1}{(\qt - \kt)^2} \right]_+,
\end{align}
where $M$ is a still unspecified reference scale. The  coefficients take the following form
\begin{widetext}
\begin{align}
  \label{eq:J_fordistributions_KT}
\hat{C}_{gg*}^{(1)f, k_T}&(x_0, \zeta_B; \qt, \kt, \mu)  = \frac{\alpha_s C_A}{2 \pi} \Bigg\{ \delta^{(2)}(\lt) \bigg[
 \ln \frac{\qt^2}{\mu^2} 
\left( \ln \frac{\zeta_B}{\mu^2} - \frac{\beta_0}{2 C_A} \right)
 - \frac{1}{2} \ln^2 \frac{\qt^2}{\mu^2}
%\notag \\
%&
  + \frac{67}{18} - \frac{5 n_f}{9 C_A} - \frac{\pi^2}{3} \bigg]
 \notag \\
&
+ 
 \ln \left(\frac{\zeta_B}{x_0 \qt^2} \right) \frac{1}{\pi^{}}\left[\frac{1}{\lt^2} \right]_+  
 +
\frac{1}{\pi^{}} \int\limits_0^1 dz \Bigg[
\frac{z(1-z) (\lt^2 - \qt^2)^2}{\left[z \lt^2 + (1-z) \qt^2\right]^2 \kt^2}  
%\notag \\
%&
%\hspace{2cm}
+ \frac{1}{ \lt^2} \frac{\kt^2 - 3(\qt - \kt)^2 - \qt^2}{\left[z (\qt - \kt)^2 + (1-z) \qt^2 \right] } \Bigg]
\Bigg\} , \\
 \label{eq:J_fordistributions_h_regKT}
\hat{C}^{(1)h, k_T}_{gg*}&(x_0, \zeta_B; \qt, \kt, \mu)  = \frac{\alpha_s C_A}{2 \pi } \Bigg\{ 
\delta^{(2)}(\lt) \bigg[ \ln \frac{\qt^2}{\mu^2} \left[ \ln \frac{\zeta_B}{\mu^2} - \frac{\beta_0}{2 C_A} \right] - \frac{1}{2} \ln^2 \frac{\qt^2}{\mu^2}
%\notag \\
%&
+  \frac{67}{18} - \frac{5n_f}{9 C_A} - \frac{\pi^2}{3}
 \bigg] 
\notag \\
& %\hspace{-1.4cm}
 +  
\ln \left( \frac{\zeta_B}{x_0\qt^2}\right)  \frac{1}{\pi^{}}\left[\frac{1}{\lt^2} \right]_+  
+ 
             \ln \left(\frac{\zeta_B}{\lt^2} \right)\frac{2 \left( (\lt \cdot \qt)^2 - \lt^2 \qt^2 \right)}{\pi^{}\lt^2 \qt^2 \kt^2}% \notag \\
%& \hspace{2cm} 
+ 
\frac{1}{\pi^{}} \int\limits_0^1 dz \frac{(\qt^2 - \lt^2)2 \kt \cdot \lt }{\left[z \lt^2 + (1-z) \qt^2 \right] \lt^2 \kt^2} \Bigg\},
\end{align}  
\end{widetext}
with $\lt = \kt - \qt$.
Even though both $\eta_a$ and the function $\bar{f}^{(1)}$ depend within this scheme on a certain reference scale $M$, the dependence on this scale cancels between both contributions, and we remain only with the parameter $x_0$ which is of order one. The scale $M$ remains therefore unspecified and can be used to satisfy the ordering condition Eq.~\eqref{eq:ordering}. A possible and suitable  choice is then  $\zeta_B^i =C \cdot \qt^2$, with $C$ another constant of order one, which eliminates a potential large logarithm in the coefficients and  which specifies eventually  $M = |\qt|$. The complete resummed gluon TMDs take then the following final form
\begin{widetext}
\begin{align}
  \label{eq:kT_def_Coeff}
    f^{k_T}_{g}(x, \zeta_{B,f}, \qt, \mu_f) & =  \exp \left\{\int\limits_{\mu_i}^{\mu_f} \frac{d \bar{\mu}}{\bar{\mu}} \gamma_G\left(\alpha_s(\bar{\mu}), \ln \frac{\zeta_{B,f}}{\bar{\mu}^2} \right) \right\} \cdot  \int d^2 \qt' G_{\text{CS}}\left(\frac{1}{2}\ln \frac{\zeta_{B,f}}{M^2}, \qt, \qt', \mu_i \right) 
\notag \\
&
\hspace{1cm}
 \cdot \int \frac{d^2 \kt}{\pi}  C^{f,k_T}_{gg^*} \left(x_0, M^2, \qt', \kt, \mu_i\right) \cdot \mathcal{G}^{k_T}\left(\ln \frac{x_0}{x}, \ln \frac{M}{p_B^-}, \kt, \bar{f}^{(1)}_{k_T}\right), \notag \\
 h_{g}^{k_T}(x, \zeta_B, \qt, \mu_f) & =  \exp \left\{\int\limits_{\mu_i}^{\mu_f} \frac{d \bar{\mu}}{\bar{\mu}} \gamma_G\left(\alpha_s(\bar{\mu}), \ln \frac{\zeta_{B,f}}{\bar{\mu}^2} \right) \right\} \cdot  \int d^2 \qt' G_{\text{CS}}\left(\frac{1}{2}\ln \frac{\zeta_B^f}{M^2}, \qt, \qt', \mu_i \right) 
\notag \\
&
\hspace{1cm}
 \int \frac{d^2 \kt}{\pi}  C^{h,k_T}_{gg^*} \left(x_0, M^2, \qt', \kt, \mu\right)  \mathcal{G}^{k_T}\left(\ln \frac{x_0}{x}, \ln \frac{M}{p_B^-}, \kt, \bar{f}^{(1)}_{k_T}\right),
\end{align}  
\end{widetext}
where $\zeta_B^i = M^2$ and $\mu_i$ are yet unspecified scales. Reading the above expressions from the right to the left, one first evolves the unintegrated gluon density through BFKL evolution up to the hadron momentum fraction $x$, with a corresponding factorization uncertainty parametrized through $x_0$. The unintegrated gluon distribution is then convoluted with the NLO TMD coefficient in transverse momentum $\kt$. The resulting expression defines then the gluon TMD with transverse momentum $\qt'$ at a scale $\zeta_B^i = M^2$ and renormalization point $\mu_i$.  While $M$ is an arbitary scale, introduced to define the $k_T$-factorization scheme, it is naturally chosen to be of the order of the transverse momentum $\qt'$; the same is true for the choice of the renormalization point $\mu_i$. In a next step it is therefore needed to evolve this gluon TMD  both in $\zeta$ (rapidity evolution of soft gluons) and finally in the renormalization scale to its final values, where rapidity evolution of solution gives rise to a convolution in transverse momentum $\qt '$. The above expressions for CSS evolution (combined evolution in $\zeta$ and $\mu$)  are the conventional expresses found in the literature, while they are expressed in transverse momentum instead of transverse coordinate space.  In particular,  the CS Green's function in transverse momentum space is  obtained from the frequently used transverse coordinate expression through 
\begin{align}
  \label{eq:GCS1loop}
  G_{CS}(\Delta y, \qt, \qt', \mu) & = \int \frac{d^2 \bt }{(2 \pi)^2} e^{- i \bt \cdot (\qt_t - \qt')} e^{\Delta y \cdot \tilde{K}_{CS}(\bt, \mu)}.
\end{align}
At 1-loop, $\tilde{K}_{CS}$  is given by Eq.~\eqref{eq:defCS}. For perturbative higher orders see  {\it e.g.}   Sec.~III of \cite{Echevarria:2015uaa}, where it is needed to include a relative factor of 2 with respect to the convention employed in this paper. Apart from perturbative higher order corrections, one might also consider RG evolution from the scale $\mu_i$ to a suitable renormalization point of the CS Green's function; finally it is also possible to include a model for non-perturbative effects. If one restricts oneself on the other hand to the leading order kernel Eq.~\eqref{eq:defCS}, the above integral can be easily evaluated and one finds
\begin{align}
  \label{eq:CSGreen_LL}
  G_{\text{CS}}^{\text{LL}} (\Delta y, \qt_1, \qt_2, \mu)  =& \frac{\Gamma(1- \overline{\alpha}_s \Delta y)}{(\qt_1 - \qt_2)^2\Gamma(\overline{\alpha}_s \Delta y)} \notag \\ & \times
\left(\frac{(\qt_1 - \qt_2)^2 e^{- 2 \gamma_E}}{\mu^2} \right)^{\overline{\alpha}_s \Delta y},
\end{align}
with $\overline{\alpha}_s ={\alpha_s C_A}/{\pi}$. Note that convergence of the Fourier integral requires $\overline{\alpha}_s \Delta y < 1$ for the above expression. While the kernels of CSS and BFKL evolution in Eq.~\eqref{eq:kT_def_Coeff} are well known, our result provides as a new element the perturbative coefficient which connects both evolution equations up to NLO accuracy. In particular, a complete next-to-leading logaritmic resummation of both BFKL and CSS logarithms requires to combine the NLO coefficient Eqs.~\eqref{eq:J_fordistributions_KT}, \eqref{eq:J_fordistributions_h_regKT} with the unintegratd gluon distribution evolved with the NLO BFKL kernel  \cite{Fadin:1998py} as well as CSS evolution with  NLO anomalous dimension Eq.~\eqref{eq:anom}, and the corresponding expression for the CS kernel,  see \cite{Echevarria:2015uaa} for a compact summary up to NNLO accuracy  of these elements in the gluonic channel. It would be very interesting to compare this result to the low $x$ expansion of exact  N$^{3}$LO results for the gluon TMD PDFs \cite{Luo:2020epw}. From a technical point of view, this would require to construct a partonic unintegrated gluon distribtion, following   Eq.~\eqref{eq:ugd_general}, but using  NLO quark and gluon impact factors.\\

%%%%%%%
While the identification of the evolution parameter according to the  $k_T$-scheme, as used  above,  provides a direct generalization of the
collinear result and is been often employed in fits of the
unintegrated gluon density, see {\it e.g.} \cite{Hentschinski:2012kr,
  Hentschinski:2013id, Chachamis:2015ona,Celiberto:2019slj}, it is not
necessarily the most adequate to describe rapidity evolution of the
system. An alternative form would be to identify $\eta_a$ with the
maximal rapidity of the soft gluonic system, $\eta_a = y_c$ or with
the rapidity of the hard final state, {\it i.e.} the Higgs boson
$\eta_a = y_H$. While $\eta_a = y_c$ eliminates entirely the need for
CS evolution, it also ignores the rapidity of the hard event (Higgs
boson) in the energy evolution of the TMD PDF. The choice appears
therefore to be possilbe, but inadequate. The choice $\eta_a = y_H$
evolves the unintegrated gluon density through BFKL evolution up to
the rapidity of the hard even, while CS evolution addresses the
possible differences $y_c - y_H$, where both $y_c > y_H$ and
$y_c < y_H$ is possible. Introducing furthermore the parameter
$\delta y$, which allows  to address the scale uncertainty
associated with high energy factorization, {\it i.e.} high energy
evolution describes dependence on $\bar{y}_H = y_H + \delta y$ and
$e^{\pm \delta y}$ is taken to be of order one, one finds 
\begin{widetext}
\begin{align}
  \label{eq:diff}
   f^{\text{rap.}}_{g}\left(\bar{y}_H, y_c, \qt, \mu \right) & =  \exp \left\{\int\limits_{\mu_i}^{\mu_f} \frac{d \bar{\mu}}{\bar{\mu}} \gamma_G\left(\alpha_s(\bar{\mu}), \ln \frac{\zeta_{B,f}}{\bar{\mu}^2} \right) \right\} \cdot
 \int d^2 \qt'G_{\text{CS}}\left(y_c-\bar{y}_H, \qt, \qt', \mu_i\right)
\notag \\
&
\hspace{1cm} \cdot
            \int \frac{d^2 \kt}{\pi}  C^{f}_{gg^*} (\zeta_B^i, y_c, \bar{y}_H, \qt, \kt, \mu, 0) \cdot  \mathcal{G}\left(\bar{y}_H - y_0, y_0, \kt, 0\right),
\notag \\
 h^{\text{rap.}}_{g}\left(\bar{y}_H, y_c, \qt, \mu \right) & =  \exp \left\{\int\limits_{\mu_i}^{\mu_f} \frac{d \bar{\mu}}{\bar{\mu}} \gamma_G\left(\alpha_s(\bar{\mu}), \ln \frac{\zeta_{B,f}}{\bar{\mu}^2} \right) \right\} \cdot
 \int d^2 \qt'G_{\text{CS}}\left(y_c-\bar{y}_H, \qt, \qt', \mu_i\right)
\notag \\
&
\hspace{1cm} \cdot
            \int \frac{d^2 \kt}{\pi}  C^{h}_{gg^*} (\zeta_B^i, y_c, \bar{y}_H, \qt, \kt, \mu, 0) \cdot  \mathcal{G}\left(\bar{y}_H - y_0, y_0, \kt, 0\right),
\end{align}  
\end{widetext}
where $\zeta_B^i =  (M_H^2 + \qt^2)e^{-2 \delta y}$,  $\zeta_B^f =  (M_H^2 + \qt^2)e^{2(\bar{y}_H-y_c )}$ and $y_0$ is a parameter of the order of the hadron rapidity. Note that within this frame, the TMD PDFs no longer depend on the hadron momentum fraction, but rather on rapidity. While this might appear strange at first sight, it is natural from the point of view of high energy factorization where the momentum fraction is -- in the case of the $k_T$-factorization scheme --  merely an evolution parameter fixed through the kinematics of the final state, while the above rapidity scheme uses a  different choice for this evolution parameter.

\subsection{Relation to previous results in the literature}
\label{sec:previous}

Before we conclude, we briefly discuss the relation of the above
results to results in the literature. In principle the TMD gluon
distribution has been already study within high energy factorization
at NLO in \cite{Xiao:2017yya}, previous studies with a similar scope
are \cite{Zhou:2016tfe} and \cite{Mueller:2012uf,Mueller:2013wwa}. At
the level of real corrections, the contributions seem to be identical
at the level of Feynman diagrams, leaving aside the absence of the
multiple reggeized gluon exchange in the current discussion. The set
of virtual diagrams of \cite{Xiao:2017yya} is on the other hand
clearly reduced with respect to the ones considered in this work,
Fig.~\ref{fig:virtual2}. Indeed, the authors of \cite{Xiao:2017yya}
seem to consider only self-energy corrections to the Wilson line in their
approach. This difference is direcly related to the fact that
\cite{Xiao:2017yya} makes use of the so-called `shock-wave picture'
for the calculation of next-to-leading order corrections. While this
is a frequently used frame for the calculations within the
CGC-framework at both LO and NLO, it does not allow to recover the term proportional to
$\beta_0$ in the anomalous dimension Eq.~\eqref{eq:anom}, as already
noted by the authors of \cite{Xiao:2017yya}. We believe that this
constitutes an important advantage of the framework of the high energy
effective action, since it does not only allow to recover the
double-logarithmic contribution to CSS resummation, ({\it i.e.} the
Sudakov form factor  of \cite{Xiao:2017yya}), but also its
single logarithmic terms.
\\

Another point in which we somehow differ with \cite{Xiao:2017yya}, see
also the discussion in \cite{Zhou:2016tfe}, is the statement that BFKL
and CS evolution cover by default distinct regions of phase space. As
outlined above, this problem does strictly speaking not occur, if the
evolution variable of the gluon density is identified with the hadron
momentum fraction. To clarify this point further and to put this into
context with the discussion in \cite{Xiao:2017yya}, we consider in the
following the combined $z\to 0$ and $z \to 1$ singularities of the real
corrections of our 1-loop result. Since the treatment is slighly more
involved for the linearly polarized gluon TMD,
Eq.~\eqref{eq:splitting_fordistributions_h}, we focus in the following
on the unpolarized case,
Eq.~\eqref{eq:splitting_fordistributions_f}. Removing regulators
through taken the limits $\rho, \sigma \to \infty$, and keeping only
singular terms, we find that the TMD splitting function reduces to
\begin{align}
  \label{eq:TMDsplitlilmit}
 \int \limits_0^1& dz  \frac{1}{(\qt - \kt)^2} \tilde{P}^{(0)f}_{gg, r} (z, \qt \kt)   \notag \\ &
 \simeq  \frac{\alpha_s C_A}{ \pi \mu^{2 \epsilon} \Gamma(1- \epsilon)}  \frac{1}{(\qt - \kt)^2}
 \int \limits_0^1 dz \left(\frac{1}{z} + \frac{1}{1-z} \right).
\end{align}
At first sight, the poles at $z=0$ and $z=1$ are therefore indeed well separated. With the rapidity of the produced gluon equal to $\eta_l = \ln \frac{|\qt - \kt| z}{(1-z) q^-}$, the above integral can be however rewritten as 
\begin{align}
  \label{eq:TMDsplitlilmit2}
 \int \limits_0^1 & dz  \frac{1}{(\qt - \kt)^2} \tilde{P}^{(0)f}_{gg, r} (z, \qt \kt) \notag \\ & \simeq 
\frac{\alpha_s C_A}{ \pi \mu^{2 \epsilon} \Gamma(1- \epsilon)}  \frac{1}{(\qt - \kt)^2}\lim_{\sigma, \rho \to \infty} \int \limits_{- \rho/2}^{\sigma/2} d \eta_l,
\end{align}
where we re-inserted the previously removed regulators $\rho, \sigma$ as cut-offs on the rapidity integral. As for the real part of the 1-loop BFKL kernel within the high energy effective action, see \cite{Hentschinski:2011tz,Chachamis:2012cc,Hentschinski:2020rfx} for an explicit construction, the above expression is proportional to an integral which extends over the entire range of rapidity. In contrast to the derivation of the BFKL kernel, the above integral is however split up  into a `soft' and a `hard' part,
\begin{align}
  \label{eq:softandhard}
   \int \limits_{- \rho/2}^{\sigma/2} d \eta_l & =  \int \limits_{- \rho/2}^{\eta_a} d \eta_l +  \int \limits_{\eta_a}^{\sigma/2} d \eta_l \notag \\  & = \left(\frac{\sigma}{2} - \eta_a\right) + \left(\eta_a + \frac{\rho}{2} \right),
\end{align}
where the distinction into `soft' and `hard' is essentially achieved through the virtual corrections. As a consequence -- while we somehow agree with \cite{Xiao:2017yya} that both pieces are well separated -- care is needed to avoid overlap between both contributions. In particular, a gluon with a certain fixed rapidity may be either counted as soft or hard but never as both. At the same time, `gaps' in rapidity should be avoided for a consistent and correct description.

\section{Conclusions and outlook}
\label{sec:concl}
In this paper we extended the framework  established in \cite{Hentschinski:2020tbi} for next-to-leading order corrections within Lipatov's high energy effective action to the case where the transition function contains an additional finite contributions. Using this extension we were able to address the special, but important case of impact factors which possess a strong hierarchy with respect to their transverse scales, as it is the case within the $k_T$-factorization setup. The latter  is characterized by a impact factor with a hard scale, {\it i.e.} the off-shell partonic coefficient, and a hadronic impact factor, characterized by transverse momenta in the non-perturbative domain. The resulting expression have been found to agree with existing results in the literature, which have been established through a study of multi particle production amplitudes in the (Quasi-)Multi-Regge Kinematics \cite{Bartels:2006hg}. Establishing this formalism at NLO within the high energy effective action is the first key result of this paper. \\

Another key result is the determination of the next-to-leading order corrections to the gluon TMD PDFs in high energy factorization, making use of the established formalism for the renormalization of matrix elements of reggeized gluon fields. While unsubtracted NLO result is subject to both rapidity divergences due to high energy factorization and rapidity divergences due to definition of the TMD PDF, the subtracted and renormalized coefficient is completely free of such divergencies. In particular we stress that  rapidity divergences related to the definition of the TMD PDF  can be treated using the soft factor,  established  within a setup based on collinear factorization. The same observation applies to the treatment of  ultraviolet divergences and their renormalization. While this behavior was to be expected and indeed constitutes a necessary requirement, it provides a non-trivial check on the correctness of our result. While we confirm earlier results in the literature which state that unpolarized and linearly polarized gluon TMD agree with each other in the dilute {\it i.e.} BFKL limit, we find that both distributions differ at NLO, which is directly related to the non-trivial tensor structure of the real NLO corrections. \\

As a next step we clarified further the relation between  BFKL and CS evolution and clarified that they describe both evolution of the  system in rapidity, {\it i.e.} BFKL of the cross-section and therefore directly related to the hard final state, while CS evolution  rapidity evolution of the soft system. Both evolution equation are therefore not independent and care is needed to avoid over-counting. Unlike previous calculations based on the CGC framework, our study further enabled us to recover the finite term proportional to $\beta_0$ in the anomalous dimension of the TMD PDFs.\\

 As an important side result of our study we find that the real NLO contribution of the unpolarized gluon TMD yield precisely the off-shell TMD gluon-to-gluon splitting function, determined in \cite{Hentschinski:2017ayz}. While \cite{Hentschinski:2017ayz} determined this splitting function from a diagrammatic approach -- essentially requiring simultaneous fulfillment of collinear and high energy limit while imposing gauge invariant production vertices -- the current study obtains the same result from the QCD operator definition of gluon TMDs. It therefore establishes an important link between both frameworks, which will be of importance to continue with these efforts. In particular the current study provides a possibility to finally determine the still missing virtual corrections to these splitting kernels, and to  formulate corresponding evolution equations. \\ 

Apart from these efforts, future studies should investigate phenomenological consequences of the derived results, which now allow to use the complete CSS resummation formulation to resum double and single logarithmic contributions with the methods of the renormalization group, extending previous results in the literature such as \cite{Mueller:2012uf,Mueller:2013wwa, Xiao:2017yya, Stasto:2018rci}. Another direction of research should address the inclusion of high density effects, along the lines of \cite{Xiao:2017yya}, but including the complete treatment of factorization scheme dependence established in this paper, asd well as contributions due to quarks.

\section*{Acknowledgments}
I would like to thank Aleksander Kusina, Krzysztof Kutak, and Mirko
Serino for collaboration at an early stage of this project. I am also
grateful to Krzysztof Kutak for his comments on the draft. Support by
Consejo Nacional de Ciencia y Tecnolog\'ia grant number A1 S-43940
(CONACYT-SEP Ciencias B\'asicas) is gratefully acknowledged.

\end{document}